\AtBeginDocument{%
  \setlength{\abovedisplayskip}{4pt plus 1pt minus 1pt}%
  \setlength{\belowdisplayskip}{4pt plus 1pt minus 1pt}%
  \setlength{\abovedisplayshortskip}{2pt}%
  \setlength{\belowdisplayshortskip}{2pt}%
}
\documentclass[lettersize,journal]{IEEEtran}

\usepackage{amsmath,amsfonts}
\usepackage{algorithm}
\usepackage{algpseudocode}
\usepackage{array}
\usepackage{textcomp}
\usepackage{stfloats}
\usepackage{url}
\usepackage{verbatim}
\usepackage{cite}
\usepackage{booktabs}
\usepackage{tabularx}
\usepackage{amsthm}
\allowdisplaybreaks
\usepackage{amssymb}

\usepackage[english]{babel}
\usepackage{graphicx}
\usepackage{dsfont}
\usepackage{etoolbox}
\AtBeginEnvironment{proof}{\setlength{\topsep}{2pt}}
\usepackage{subcaption}
\usepackage{hyperref}
\usepackage{cleveref}
\usepackage{algpseudocode}

\DeclareMathOperator*{\argmin}{argmin}
\newtheoremstyle{compact}
  {3pt}       
  {3pt}       
  {\itshape}  
  {}          
  {\bfseries} 
  {.}         
  {.5em}      
  {}          
\theoremstyle{compact}
\newtheorem{assumption}{Assumption}
\newtheorem{theorem}{Theorem}
\newtheorem{corollary}{Corollary}

\newtheorem{proposition}{Proposition}
\theoremstyle{remark}

\Crefname{assumption}{Assumption}{Assumptions}
\Crefname{assumption}{Assumption}{Assumptions}
\begin{document}
\bstctlcite{IEEEexample:BSTcontrol}
\title{Information Value-Driven Neurosymbolic Scheduling for Wireless Physical AI Inference}

\author{
    \IEEEauthorblockN{Nitisha Singh\IEEEauthorrefmark{1}, Walid Saad\IEEEauthorrefmark{1}, and Pascal Adjakpl\'e\IEEEauthorrefmark{2}\thanks{This work was supported by InterDigital.}} \\
    \IEEEauthorblockA{\small \IEEEauthorrefmark{1}Bradley Department of Electrical and Computer Engineering, Virginia Tech, Alexandria, VA, 22305, USA} \\
    \IEEEauthorblockA{\small \IEEEauthorrefmark{2}InterDigital, USA} \\
    \IEEEauthorblockA{Emails: \{nitishas, walids\}@vt.edu, pascal.adjakple@interdigital.com}
}
{


\maketitle

\begin{abstract}
Physical artificial intelligence (AI) is a key use case for emerging 6G networks, in which distributed agents, such as robots and connected vehicles, transmit their sensor observations over the uplink to perform inference for a physical task at a remote server. While task-oriented communications can help improve communication efficiency for such remote inference tasks of physical AI, existing schedulers typically allocate uplink resources based on metrics such as channel rate or age of information and do not account for an agent's contribution to the inference task. In this paper, a novel information-value (IV)-driven scheduling framework is proposed for the remote inference tasks of physical AI systems over a bandwidth-constrained uplink. A local IV metric is defined to estimate at each agent the contribution of its observation before transmission, and a global IV metric based on the Shapley value is defined at the base station (BS) to capture the contribution of each agent relative to other agents. To compute the global IV in real time, an online stochastic approximation with $O(N)$ complexity is proposed, and finite-time convergence guarantees are derived. Uplink channel resources are then allocated by minimizing a neurosymbolic energy function, in which the IV metrics are combined with channel and resource through logical rules. Experiments for a cooperative-perception task using a CARLA-based simulation pipeline with ray-traced wireless channel modeling, across two towns and diverse weather/lighting conditions, show that the proposed NeSyEBM scheduler can achieve around $15\%$ higher detection accuracy than a MaxRate scheduler while using $50\%$ less bandwidth.
\end{abstract}

\vspace{-2mm}
\section{Introduction}
\IEEEPARstart{N}ext-generation wireless cellular systems, such as 6G and beyond, must support emerging services that include cyber-physical systems, extended reality, and physical artificial intelligence (AI) \cite{6gusecases}. Central to these emerging applications is the concept of remote inference in which distributed edge devices (e.g., sensors, drones, or robots) continuously collect high-dimensional data and offload it to edge servers equipped with large AI models, e.g., multimodal perception or foundation models, whose computational and memory requirements exceed the capabilities of the devices\cite{6gedgeai}. Since these devices are mobile and the server often combines the observations of several devices, the data is transmitted over the wireless uplink. However, AI-based inference models require large amounts of data for a specific task (e.g., classification, detection); a perfect bit-by-bit transmission of the payload is often unnecessary and can waste the network's limited spectrum and energy resources. Instead, a promising approach to improve communication efficiency is to extract and transmit only the task-relevant features \cite{9606667}. For instance, in an autonomous driving application that intends to detect pedestrians from images, transmitting the pixels that represent the sky is unnecessary. Transmitting only the task-relevant features requires a shift from traditional bit transmission to \emph{task-oriented (or goal-oriented) communications}. In a goal-oriented network, the primary objective is no longer the perfect reconstruction of every transmitted bit, but rather the successful and timely execution of a specific downstream task for the served application \cite{11563815}. Physical AI inference applications involve large volumes of uplink multi-modal, high-dimensional data traffic (e.g., LiDAR, high-resolution video), and have strict latency deadlines \cite{11510260}. Since the wireless uplink has a constrained bandwidth, optimizing for traditional quality-of-service (QoS) metrics cannot guarantee optimal task performance for these applications.  
\par To overcome the limitations of traditional QoS metrics, the wireless network must be able to understand and exploit the information value (IV) of the data streams it carries. In the context of remote inference, IV represents the meaning, utility, and relevance of a specific data sample to the accuracy of the downstream task for the served application. 
By quantifying the IV, the network can intelligently prioritize the transmission of only the most critical task-relevant features. If done properly, such task-oriented network management can help conserve scarce uplink resources while maximizing the expected inference gain. 
\par Quantifying IV is also a key enabler of semantic communications (SC), in which the wireless network has knowledge-driven physical and network layers that understand the needs of the user and the application~\cite{lesdata}. This knowledge allows the network to quantify the IV of the data it carries and utilize it in resource management functions, such as scheduling. However, the notion of IV is inherently application-specific and context-dependent. Translating IV into a practical, real-time scheduling policy for physical AI also presents a number of challenges.
First, the data-valuation metrics for IV that are used in the field of machine learning (ML) are typically evaluated post-inference and are tied to the specific model (particular to the inference application) \cite{ghorbani2019data}. This is not applicable in wireless resource allocation because the scheduler must quantify the IV a priori, before allocating resources, and therefore, before the inference task is actually executed. Second, even if these post-inference data valuation metrics can be adapted, calculating them requires extensive computational overhead (e.g., combinatorial evaluations), which makes them impractical for real-time scheduling. Finally, an IV-aware scheduler must integrate the IV metric alongside dynamic, task-dependent constraints and channel conditions. Therefore,  there is a need for a wireless resource allocation framework that allows IV to be defined and adapted according to the specific task requirements and dynamic system context.
\vspace{-3mm}
\subsection{Related Works}
Recent literature on resource allocation for remote inference has primarily relied on age of information (AoI) to design scheduling policies under bandwidth constraints \cite{10559951, 11206222, aoiedgehierarchy, 10620407}. Meanwhile, a number of works such as \cite{10.1145/3302509.3311050, valuematters, lei2025learningvalueinformationjoint} have adapted the traditional value of information (VoI) for specific system goals, such as networked control and vehicle-to-everything (V2X) communications. While these metrics are often used as proxies for data utility, they fail to capture the value for the inference application. For instance, in an autonomous control scenario, the value of a sensor depends heavily on the context. In low-visibility conditions (e.g., at night), LiDAR features are more valuable because they offer higher reliability for depth inference than camera features. An AoI- or VoI-based scheduler may not be able to prioritize the LiDAR stream, as they optimize for age or estimation error.
\par Research on goal-oriented SC systems focuses on optimizing transmissions for the downstream task. Deep learning (DL) based SC has been studied in remote inference systems \cite{colseminf, attnsemcolinf, llcinfsem}. The authors in \cite{colseminf} leverage joint source and channel coding (JSCC) for collaborative edge inference. In \cite{attnsemcolinf}, a transformer-based SC system is proposed to select only the task-relevant patches for image transmission. In \cite{llcinfsem}, a DL-based SC system is developed for optimizing latency in multi-state edge inference. Thus, research in goal-oriented SC \cite{colseminf, attnsemcolinf, llcinfsem} optimizes what each device encodes and transmits, but they do not address the uplink resource allocation problem based on the value of the data to the task. 
\par While the work in \cite{semcomdata} highlights the need for a semantics-based metric for random access in SC systems, the discussion is limited to freshness, and the definition of value is limited to a cost versus benefit tradeoff. The authors in \cite{10494374} propose a metric for semantic value; however, its application is limited to text transmission. Such metrics are unsuited for multi-agent physical AI inference tasks, e.g., cooperative perception, where the value of an agent's observations often depends on what other agents are observing simultaneously. Semantic and task-oriented communications have been applied to cooperative perception among connected vehicles. In \cite{11355867}, the authors propose an importance-aware feature-selection network to transmit the relevant semantics for a cooperative perception task. Resource allocation for cooperative perception has been studied under computing-efficiency and delay objectives in \cite{10413956}. In \cite{11106181}, a DDQN-based scheduling framework for cooperative perception is proposed that uses channel state and perceptual semantics as input. 
However, these schedulers do not account for redundancy between the data of scheduled agents, e.g., two vehicles that observe the same occluded target. In summary, the prior art on goal-oriented SC and remote inference \cite{10559951, 11206222, aoiedgehierarchy, 10620407, 10.1145/3302509.3311050, valuematters, lei2025learningvalueinformationjoint, colseminf, attnsemcolinf,llcinfsem, semcomdata, 10494374, 11355867, 10413956, 11106181} has two key limitations. First, there is no IV metric that quantifies how much an agent's data improves the inference, accounting for both its observations and the observations of the other agents. Without such a metric, the network cannot distinguish redundant or complementary observations and schedule agents that add little to the inference. Second, estimating such a metric for real-time scheduling is challenging. Computing the exact value of an agent relative to the other agents requires the inference outcomes of all $2^N$ subsets of agents, but the network only observes the inference outcomes of the subset of agents it schedules. Hence, the value estimate is biased by the network's scheduling decisions. Third, the scheduling criterion must be adaptable to different systems, as fixed-form network objectives, e.g., rate- or AoI- based, must be redesigned when the task changes and cannot express conditional requirements that couple the IV, the channel, and the resources. Learned black-box policies capture such requirements implicitly; however, they must be retrained when task conditions change, and they generalize poorly under domain shift. 
\par In our prior work~\cite{singh2026globecom}, we quantified IV using the Shapley value concept and proposed a low-complexity algorithm to update the IV in real time. However, the work in \cite{singh2026globecom} did not provide guarantees on the accuracy of the online IV estimate or on the bias of this estimate due to scheduling decisions. Moreover, the scheduler in \cite{singh2026globecom} maximized a fixed-form objective, namely the IV-weighted transmission success. 
\vspace{-5mm}

\subsection{Contributions}
The main contribution of this paper is a novel framework to quantify IV and to leverage it along with channel, resource, and logical constraints for optimal resource allocation to improve the inference performance of the remote inference tasks of physical AI agents when they are deployed over a resource-constrained wireless network. In particular, we consider a wireless network that must support, in the uplink, a group of distributed agents engaged in a physical AI task. We propose a framework to quantify IV both at the agent, from its own observation, and at the BS relative to the observations of the other agents. We then propose a context-aware IV-based objective function that encodes the logical relationships between channel, IV, and resource constraints, which the BS optimizes to allocate RBs. Our key contributions include:
\begin{itemize}
    \item We propose a local IV metric that estimates, from the agent's observation, the marginal utility that observation would contribute to the inference task if transmitted without access to the global context.  Meanwhile, at the BS, we introduce a global IV metric that captures the importance of an agent's observations relative to the observations from other agents, via the concept of Shapley value from cooperative game theory.
    \item Since calculating the Shapley value entails significant computational overhead, we propose an online stochastic gradient descent (SGD) based approximation to estimate the global IV in real time with $O(N)$ complexity. We establish finite-time convergence guarantees for this online estimator under both fixed and time-varying scheduling. We also prove that $\epsilon$-greedy exploration is necessary for convergence. For the network, these results characterize the accuracy of the BS's estimate of global IV and how quickly it can converge to a close estimate of the true global IV. We validate these guarantees directly against real data from a cooperative-perception task.
    \item To incorporate context-awareness into the resource allocation, we propose an IV-based objective function that exploits the logical dependencies between different attributes relevant for decision-making (global IV, channel states, application performance) via the neurosymbolic energy-based model (NeSyEBM) \cite{10.24963/ijcai.2023/461}. The BS minimizes the NeSyEBM energy, which is a weighted sum of the penalty for violating constraints formulated as a convex relaxation of symbolic logical rules. We propose a solution based on unsupervised stochastic policy optimization to learn the NeSyEBM scheduler.
    \item  We validate the proposed NeSyEBM scheduler on a realistic cooperative perception system, with heterogeneous sensing agents, using CARLA and ray-traced wireless channels in two CARLA towns. Simulation results show that the NeSyEBM scheduler improves the detection accuracy by around $15\%$ over MaxRate and up to $7.7\%$ over the VoI scheduler, while using approximately $50\%$ less bandwidth than both, and achieving within $2\%$ of the maximum possible accuracy. Ablation studies show that removing local IV reduces the cooperative gain per MHz by up to $62.5\%$ and removing the global IV by up to $50\%$, showing that IV metrics improve efficiency by reducing redundant transmissions.
\end{itemize}
\par The rest of this paper is organized as follows. Section II presents the system model and the definition of IV. Section III describes the proposed NeSyEBM-based scheduler and the proposed solution to the resource allocation optimization problem. Section IV discusses the simulations and results. Finally, we conclude this paper in Section V. \vspace{-2mm}
\begin{figure}
    \centering
    \includegraphics[width = \linewidth]{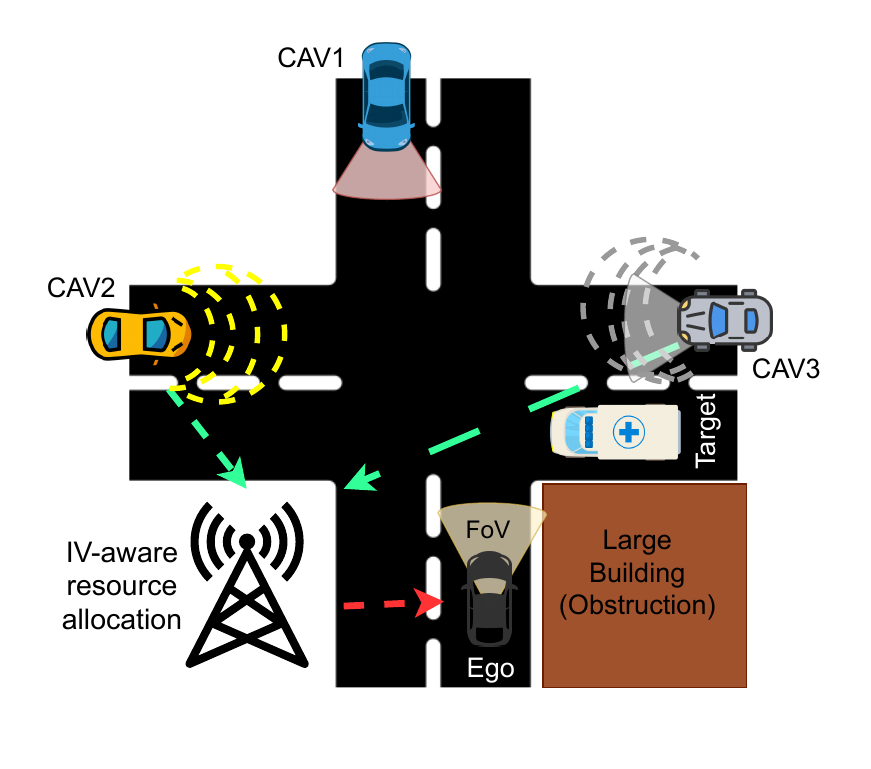}
    \caption{\small{Illustrative example of the proposed IV-aware resource allocation framework for a cooperative perception task. The framework applies to any group of physical AI agents (e.g., robots, drones, or CAVs) that perform a shared remote inference task. Here, CAVs with different sensing capabilities assist an ego vehicle, whose field of view is occluded by a large building, in detecting a target vehicle that could cause a collision. Based on their IV, the NeSyEBM scheduler allocates uplink RBs to the most informative CAVs (CAV2 and CAV3), which transmit their observations over the uplink (green), and the BS sends the detections to the ego vehicle over the downlink (red).}} \vspace{-3.8mm}
    \label{systemmodel}
\end{figure}
\section{System Model}\label{sys2}

We consider a single-cell goal-oriented uplink wireless network involving a set of distributed physical AI agents $\mathcal{N}$ and a base station (BS) operating in a time-indexed manner $t\in \{0, 1,... \}$. Each agent $n \in \mathcal{N}$ acts as an independent physical AI equipped with diverse sensors (e.g., LiDAR, radar, RGB camera). Here, these physical AI agents could be robots or connected autonomous vehicles (CAVs). At each time step $t$, agent $n$ captures a raw, high-dimensional observation $\boldsymbol{x}_n$ of its local environment and transmits it to the BS for remote inference, as shown in Fig. \ref{systemmodel}, for an illustrative CAV use case. A goal-oriented remote inference application, represented by a function $f$, is executed at the BS, requiring observations from the agents to infer a target variable $y$. For example, in Fig.~\ref{systemmodel}, $f$ is a cooperative 3D object detection model that fuses the features received from the scheduled CAVs, and $y$ is the set of bounding boxes of the surrounding vehicles, including the occluded target. The agents are mobile and physically separated, and the inference model at the BS must fuse the observations of several agents. Hence, the agents must transmit their observations over the wireless network's uplink to the BS, which hosts the inference model and serves as a central data sink. However, in practical cellular networks, the uplink bandwidth is constrained and divided into a finite set of orthogonal frequency division multiple access (OFDMA) resource blocks (RBs). Therefore, to improve efficiency in limited bandwidth, the BS must schedule only a limited number of RBs to the subset of agents whose data provides the highest utility for inferring $y$. We propose a two-fold IV-driven framework consisting of two components: a local IV and a global IV. The local IV metric captures the instantaneous task relevance of an agent's current observation before transmission, allowing each agent to quantify the semantic utility of its locally observed data. The global IV metric is used by the BS to quantify the relative importance of each agent's features to the features obtained from other agents. This enables the BS to allocate limited channel resources to the agents that yield the highest gain for the inference task. \emph{Essentially, local IV allows the physical AI agent to assess the semantic utility of its current observations, and the global IV metric allows the network to control when the transmission would occur}. Using these two metrics, the BS can approximate the value of each agent's observation before allocation, and allocate RBs to the agents whose observations complement each other rather than to agents that are individually informative or have the best channels.
\vspace{-4mm}
\subsection{Communication Model}
 The communication between the agents and the BS occurs over a shared uplink wireless channel. The BS controls the uplink wireless channel, which is divided into $M$ RBs, each with bandwidth $B$. At each time slot $t \in \{0, 1,..\}$, the BS selects an integer RB allocation $\boldsymbol{r}_t=[r_{1,t},..r_{N,t}]^{T}$, where $r_{n,t} \in \mathbb{Z}_{\geq 0}$ is the number of RBs allocated to agent $n$. The feasible allocation set is given by $\Omega = \{\boldsymbol{r}_t \in \mathbb{Z}_{\geq 0}^{N}: \sum_{n \in \mathcal{N}} r_{n,t} \leq M \}$. We define a scheduling indicator $a_{n,t} = \mathds{1}\{r_{n,t} > 0\}$, variable whereby, $a_{n,t} = 1$ means that agent $n$ is scheduled in slot $t$, while $r_{n,t}$ specifies the number of RBs assigned to that agent. 
The uplink wireless channel between the agent and the BS is modeled by a set of propagation paths $\mathcal{P}_{n,t}$. For each propagation path, $l \in \mathcal{P}_{n,t}$, the complex path coefficient $p_{l,n,t}$ captures the attenuation and the phase shift along the propagation path. The effective channel gain is given by $ h_{n,t} = \sum_{l \in \mathcal{P}_{n,t}}|p_{l,n,t}|^2$. Since the BS allocates orthogonal RBs to each agent, there is no intra-cell interference. However, we account for constant inter-cell interference from neighboring cells. The instantaneous signal-to-noise plus interference ratio (SINR) of the channel between agent $n$ and the BS at time $t$ can be given by $ \gamma_{n,t} = \tfrac{P |h_{n,t}|^2}{\sigma_{\text{noise}}^2 + I_{\text{inter}}}$, where $P$ is the transmit power, $\sigma^{2}_{\text{noise}}$ is the noise power, and $I_{\text{inter}}$ is the average out of cell interference power. When agent $n$ is allocated resources at time $t$, it transmits a payload of size $L_{n,t}$, which depends on the modalities available to the agent and the encoding. For an allocation $r_{n,t} > 0$, the achievable uplink rate is given by $R_{n,t}(r_{n,t}) = r_{n,t} B \log_2(1 + \gamma_{n,t})$. The transmission time is given by $D_n(r_{n,t}) = \tfrac{L_n}{R_{n,t}(r_{n,t})}$. Let $D_{\textrm{max}}$ be the maximum allowable transmission delay; then, a transmission is successful if $D_{n}(r_{n,t})< D_{\textrm{max}}$.
The minimum SINR required to support a successful transmission is given by $\gamma_n^{\text{req}}(r_{n,t}) = 2^{\frac{L_n}{r_{n,t}B D_{\text{max}}}} - 1.$ Equivalently, the successful transmission indicator can be expressed in terms of the SINR as, 
\begin{equation}
    o_{n,t}(r_{n,t}) = \mathds{1}\{r_{n,t} > 0, \gamma_{n,t} \geq \gamma_{n,t}^{\textrm{req}}(r_{n,t})\}.
\end{equation}
\par At the beginning of each time slot $t$, each agent $n \in \mathcal{N}$ generates one or more high-dimensional observation vectors. We define $\mathcal{M}_n$ as the number of observation modalities available to agent $n$ and $x_{n,t} = \{x_{n,t}^{(m)} : m \in \mathcal{M}_n\}$ as the observation from agent $n$, where $x_{n,t}^{(m)}$ is the observation associated with modality $m$. In bandwidth-constrained remote inference applications, the BS must decide how agents must be allocated RBs without knowing the utility of their observations. Thus, each agent extracts a score from its local observation, which we define as the local IV $l_{n,t}$, and sends it to the BS when requesting uplink resources. To formalize the local IV notion, we use $U_t(S)$ to capture the \emph{inference utility} obtained at time $t$, when the BS receives observations from a subset of agents $\mathcal{S} \subset \mathcal{N}$. The utility $U_t(\mathcal{S})$ is task-dependent and defined by the inference application. For instance, in an image classification task, $U_t(\mathcal{S})$ can be the classification accuracy or the negative cross-entropy loss of $f$ given the observations of $\mathcal{S}$. In the cooperative perception task of Fig.~\ref{systemmodel}, $U_t(\mathcal{S})$ can be the F1 score of the bounding boxes detected from the features of the CAVs in $\mathcal{S}$, or an indicator of whether the occluded target is detected. The local IV is then defined as the standalone utility of its observation as:
\begin{equation}\label{localivdef}
    u_{n,t}^{\textrm{loc}} = \frac{U_t({n}) - U_t({\emptyset})}{U_t({\mathcal{N}}) - U_t({\emptyset}) + \mu_0},
\end{equation}
where $\mu_0 > 0$ is a small constant. However, the utility in \eqref{localivdef} is not available to the BS before transmission of the local observations. Therefore, the local IV is estimated by the agent using a learned predictor and reported to the BS. Let $\boldsymbol{\xi}_{n,t}$ be a low-dimensional feature vector extracted from the current observation $x_{n,t}$ by a function $f^{\textrm{loc}}$. The local IV score is then estimated as, 
\begin{equation}
    \ell_{n,t} = g_{\eta}(\boldsymbol{\xi}_{n,t}),
\end{equation}
where $g_{\eta}$ is a scalar-valued predictor with parameters $\eta$. The predictor is trained offline by minimizing
\begin{equation}
    \eta^{*} = \argmin_{\eta}\sum_t \sum_{n \in \mathcal{N}}(g_{\eta}(\boldsymbol{\xi}_{n,t}) - u_{n,t}^{\textrm{loc}})^2.
\end{equation}
Consequently, $g_{\eta}(\boldsymbol{\xi}_{n,t})$ approximates the conditional standalone utility of agent $n$, given the feature vector $\boldsymbol{\xi}_{n,t}$, $g_{\eta}(\boldsymbol{\xi}_{n,t}) \approx \mathbb{E}[u^{\textrm{loc}}_{n,t}|\boldsymbol{\xi}_{n,t}]$. During online deployment, agent $n$ sends $l_{n,t}$ as scheduling metadata before payload transmission. Then, the BS utilizes $\ell_{n,t}$, along with other metrics, to determine the RB allocation. Thus, $\ell_{n,t}$ enables the BS to estimate the utility of agent $n$'s observations without allocating uplink resources to the agent first or observing how its data contributes to the inference. 
Let $\mathcal{S}(\boldsymbol{r}_t) = \{n \in \mathcal{N} : o_{n,t}(r_{n,t}) = 1\}$ be the set of agents whose payloads are delivered within $D_{\textrm{max}}$ under the allocation $\boldsymbol{r}_t$. The goal of the BS is to solve
\begin{equation}\label{pr}
\max_{\boldsymbol{r}_t \in \Omega} \; U_t\big(\mathcal{S}(\boldsymbol{r}_t)\big),
\end{equation}
in every slot $t$, i.e., to allocate the $M$ RBs such that the inference utility is maximized. 
Solving \eqref{pr} requires $U_t(\mathcal{S})$ for every possible subset $\mathcal{S}$ before transmission. However, $U_t$ is observed after the inference and only for the scheduled subset. Hence, before transmission, the BS must estimate the value of each agent's observation via the local IV, and its value relative to the other agents via the global IV.
\vspace{-3mm}

\subsection{Global IV metric}
The local IV characterizes the relevance of an individual agent's observation at time $t$. However, the local IV does not account for the \emph{relative importance of an agent's observations with respect to other agents}. For instance, in Fig.~\ref{systemmodel}, CAV1 and CAV3 may both capture the target in their camera views. Each observation is individually informative, but once CAV3's camera and LiDAR features are received, CAV1's camera features add little. In contrast, CAV2 may observe only part of the target and thus have a moderate local IV, yet its LiDAR depth information from a different viewpoint complements CAV3's observation. Therefore, we introduce the global IV metric to quantify the marginal task contribution of each agent relative to the other agents. 
We define the global IV in terms of the contribution of the feature to the application's performance. The accuracy/loss of the inference result can be viewed as a measure of the predictiveness or value of the feature. Let $U$ be an arbitrary measure of predictiveness (e.g., accuracy or negative loss). We adopt the Shapley value as the global IV metric, where the global importance $\psi_{n, t}$ of agent $n$ at time $t$ is defined as the average marginal gain in predictiveness across all possible subsets of features \cite{pmlr-v119-williamson20a}. The \emph{global IV} is given by,
\begin{equation}\label{shapleyeq}
    \Phi_{n,t} = \sum_{S \in \mathcal{N} \backslash \{n\}} \frac{1}{N}\left(\tfrac{N - 1}{|S|}\right)^{-1} \left(U_t(S \cup \{n\}) - U_t(S)\right).
\end{equation}
\begin{algorithm}[t!]
\caption{Online Stochastic Shapley Value Estimation}
\label{alg:online_shapley}
\begin{algorithmic}[1]
\State \textbf{Input:} Initial estimate $\boldsymbol{\psi}_0$, learning rate $\alpha$, exploration rate $\epsilon$, number of agents $N$
\For{each time slot $t = 1, 2, \dots$}
    \State \textbf{1. Schedule:}
    \State With probability $\epsilon$, draw $\mathcal{S}_t$ uniformly from the proper non-empty coalitions and allocate the minimum RBs that realize it
    \State Otherwise, schedule according to the scheduling policy
    \State \textbf{2. Observe system state:}
    \State Receive the membership vector $\boldsymbol{k}_t$ of the scheduled coalition
    \State Receive the centered utility $\bar{U}_t(\mathcal{S}_t) = U_t(\mathcal{S}_t) - U_t(\emptyset)$
    \State \textbf{3. Compute kernel weight:}
    \State $\omega(\mathcal{S}_t) \gets \dfrac{N-1}{\binom{N}{|\mathcal{S}_t|}\,|\mathcal{S}_t|\,(N-|\mathcal{S}_t|)}$
    \State \textbf{4. Compute prediction error:}
    \State $e_t \gets \bar{U}_t(\mathcal{S}_t) - \boldsymbol{k}_t^\top \boldsymbol{\psi}_{t-1}$
    \State \textbf{5. SGD update:}
    \State $\boldsymbol{\psi}_t \gets \boldsymbol{\psi}_{t-1} + \alpha\, \omega(\mathcal{S}_t)\, e_t\, \boldsymbol{k}_t$
\EndFor
\end{algorithmic}
\end{algorithm}
Here, $|\mathcal{N}| = N$. Following cooperative game theory, we refer to a subset of agents $\mathcal{S} \subseteq \mathcal{N}$ as a coalition. In our setting, a coalition is the set of agents whose observations are received by the BS in the same slot, and $U_t(\mathcal{S})$ is the inference utility achieved by this coalition. We select the Shapley value as our metric for importance based on its axiomatic properties \cite{pmlr-v119-williamson20a}. The uniqueness theorem states that the Shapley value is the only allocation that simultaneously satisfies efficiency ($\sum_n \phi_n = U_t(\mathcal{N}) - U_t(\emptyset)$) and symmetry (importance is independent of ordering), null player (redundant agents receive zero IV), and additivity (importance can compose across tasks). For our problem, the null player axiom correctly assigns zero IV to redundant agents that add no value to any coalition, the efficiency axiom allows the BS to distribute the total inference performance as IV to individual agents, and the symmetry axiom ensures that two agents making identical contributions to any coalition have the same Shapley value regardless of ordering. We assume that the BS obtains an initial estimate $\boldsymbol{\psi}_0 = [\psi_{1,0}, \dots, \psi_{N,0}]$ by evaluating the utility of several subsets of agents on historical data before deployment. This is reasonable because the BS hosts the inference model and can run such an offline calibration once. Moreover, we show later that the online update does not rely on this estimate to be accurate. From \eqref{shapleyeq}, we can see that direct computation of the Shapley value grows exponentially with the number of agents, which makes it infeasible for real-time resource allocation. To overcome this challenge, we reformulate \eqref{shapleyeq} into a weighted linear regression problem to derive the Shapley importance values\cite{3295222.3295230}. We propose a new low-complexity online approximation based on stochastic gradient descent (SGD) to track the Shapley values of agents (with $O(N)$ complexity) in real-time. Let $S_t$ be the coalition (set of agents) scheduled at time $t$, and let  $\boldsymbol{k}_t = [k_{1,t}, k_{2,t}, \ dots, k_{N,t}]^T \in \{0,1\}^N$ be the binary membership vector, i.e., $k_{n,t} =  1$ if agent $n \in S_t$, and $k_{n_t} = 0$, otherwise. For a non-empty coalition, the Shapley regression weight is defined as $\omega(S_t) = \tfrac{N- 1}{\binom{N}{|S_t|} \cdot |S_t| (N - |S_t|)}$. Let $\bar{U}(S_t)$ be the observed utility of the inference at time $t$. Then, the instantaneous squared error of the Shapley prediction model is
\begin{equation}\label{instloss}
    \mathcal{J}(\boldsymbol{\psi}) = \frac{1}{2}\cdot \omega(S_t) (\bar{U}(S_t) - \boldsymbol{k}_t^T \boldsymbol{\psi})^2,
\end{equation}
where $\psi_{t} = [\psi_{1,t}, .., \psi_{N,t}]^T$ is the online global IV estimate at time $t$. Here, $\bar{U}(S_t) \overset{\Delta}{=} U(S_t) - U(\emptyset)$, and $\tilde{U}(S_t) \approx \boldsymbol{k}_t\boldsymbol{\psi}_{t-1}$. We define the error between the estimated inference utility and the actual utility as $e_t = \bar{U}(S_t) - \tilde{U}(S_t)$. $U(\emptyset)$ represents the case in which no agents are scheduled. To update our estimate, we compute the gradient of the instantaneous loss with respect to the Shapley importance vector $\boldsymbol{\psi}$
\begin{equation}
    \nabla_{\boldsymbol{\psi}} \mathcal{J}(\boldsymbol{\psi}) = \omega(S_t) (U_t(S_t) - \boldsymbol{k}_t^T\boldsymbol{\psi}) \boldsymbol{k}_t = -\omega(S_t) \cdot e_t \cdot \boldsymbol{k}_t.
\end{equation}
Then, the Shapley values at time $t$, $\boldsymbol{\psi}_{t}:= \{\psi_{1,t}, .., \psi_{N,t}\}$ for all the features, are given as, $  \boldsymbol{\psi}_t = \boldsymbol{\psi}_{t-1} + \alpha_t \cdot \boldsymbol{k}_t \cdot \omega(S_t) \cdot e_t$,
where $\alpha_t$ is the learning rate to track the update, and its per-update complexity scales linearly with $N$. We summarize this in Algorithm \ref{alg:online_shapley}

\subsubsection{Convergence and Approximation Guarantees}
Let $q_t(S)$ be the probability that coalition $S_t = S$ is sampled at time $t$ under the current scheduling policy, and let $\boldsymbol{k}(S_t) = \boldsymbol k_t$ when $S = S_t $ define the membership vector for the coalition $S_t$, then, \vspace{-2mm}
\begin{align}
\boldsymbol{G}_t &:= \sum_S q_t(S)\omega(S)\boldsymbol{k}(S_t)\boldsymbol{k}(S)^\top, \label{eq:G-def} \\
\boldsymbol{b}_t &:= \sum_S q_t(S)\omega(S)\boldsymbol{k}(S)\bar{U}_t(S). \label{eq:b-def}
\end{align}
\eqref{eq:G-def} and \eqref{eq:b-def} define a sum over every coalition the scheduler could realize. Then, $\boldsymbol{\psi}_t^{\ast} := \boldsymbol{G}_t^{-1}\boldsymbol{b}_t$ is the minimizer of the instantaneous weighted loss \eqref{instloss} under $q_t$. Since Algorithm \ref{alg:online_shapley} approximates the true Shapley value $\boldsymbol{\Phi}_{n,t}$, the total error between the online estimate $\boldsymbol{\psi}_t$ and $\boldsymbol{\Phi}_t$ can be decomposed as,
\begin{equation}
    \lVert\boldsymbol\psi_t-\boldsymbol\Phi_t\rVert \le \underbrace{\lVert\boldsymbol\psi_t-\boldsymbol\psi_t^\ast\rVert}_{\text{tracking error}} + \underbrace{\lVert\boldsymbol\psi_t^\ast-\boldsymbol\Phi_t\rVert}_{\text{sampling bias}},
\label{eq:total-error-decomposition}
\end{equation}
The tracking error is a consequence of stochastic noise and is governed by the learning rate. The sampling bias arises because the deployed scheduler samples coalitions according to a learned policy rather than the KernelSHAP distribution ($q_{KS}$), which is known to estimate the Shapley value closely \cite{3295222.3295230}. We make the following assumptions to establish convergence guarantees.
\begin{assumption}\label{assump:A1}
    There exists $\kappa < \infty$ such that $\omega(S)\lVert\boldsymbol{k}(S)\rVert^2\le \kappa$ for all coalitions $S$ with $q_t(S) > 0$.
\end{assumption}
\begin{assumption}\label{assump:A2}
    The conditional variance of $\bar{U}_t(S)$ is bounded by some $\sigma^2 <\infty$.
\end{assumption}
\begin{assumption}\label{assump:A3}
    $\boldsymbol{G}_t \succeq \mu \boldsymbol{I}$ for some $\mu > 0$, which bounds $\lambda_{\min}(\boldsymbol{G}_t) > 0$
\end{assumption}
\begin{theorem}\label{thm:fixed-convergence}
Suppose that Assumptions \ref{assump:A1}, \ref{assump:A2}, and \ref{assump:A3} hold, $q_t \equiv q$, over $T$ slots, and that the learning rate satisfies $\alpha_t \kappa \le 1/2 \forall t$. Let $\boldsymbol{\psi}^* = \boldsymbol{G}^{-1}\boldsymbol{b}$ be the regression target induced by $q$, and let $\Lambda_{s,T} := \prod_{t=s}^{T-1}(1-\mu\alpha_t)$, with $\Lambda_{T,T} := 1$. Then,
\begin{equation}\label{eq:finite-time-bound}
\mathbb{E}\!\big[\|\boldsymbol{\psi}_T - \boldsymbol{\psi}^*\|^2\big]\! \le \!\Lambda_{0,T}\,\|\boldsymbol{\psi}_0 - \boldsymbol{\psi}^*\|^2 + \sum_{t=0}\!^{T-1} \Lambda_{t+1,T}\, 2\alpha_t^2\sigma^2.
\end{equation}
For a constant learning rate $\alpha_t \equiv \alpha$, this bound becomes $(1-\mu\alpha)^T\|\boldsymbol{\psi}_0 - \boldsymbol{\psi}^*\|^2 + \frac{2\alpha\sigma^2}{\mu}\!\big[1-(1-\mu\alpha)^T\big]$. If, $\sum_t\! \alpha_t = \infty$ and $\sum_t\! \alpha_t^2 < \infty$, then $\boldsymbol{\psi}_t \to \boldsymbol{\psi}^*$ almost surely \cite{robbins1971convergence}.
\end{theorem}
\vspace{-2mm}
\begin{proof}
    See Appendix \ref{app:proofub}
    \vspace{-2mm}
\end{proof}
\par Theorem~\ref{thm:fixed-convergence} shows that the estimation error of Algorithm~1 consists of two terms. The first term is the error of the initial estimate $\boldsymbol{\psi}_0$, which decreases by a factor of $(1-\mu\alpha)$ at every slot. The second term is caused by the randomness of the observed utilities and settles at $2\alpha\sigma^2/\mu$. Hence, the BS can use the learning rate $\alpha$ to balance a tradeoff between the convergence speed of the IV estimate and its steady-state error $2\alpha\sigma^2/\mu$. The convergence speed also depends on $\mu$, the smallest eigenvalue of $\boldsymbol{G}$. $\mu$ is large when the scheduled coalitions cover all agents evenly and small when some agents are rarely scheduled. Thus, a rarely scheduled agent slows down the convergence of the entire estimate. Theorem~\ref{thm:fixed-convergence} assumes that the BS samples coalitions from a fixed distribution. In our system, however, the scheduling policy changes over time as it reacts to the current IV estimates and channel states, and the utility of each coalition changes as the scene evolves. Since $\boldsymbol{\psi}^*_t$ depends on both the sampling distribution $q_t$ and the utilities, the value that Algorithm \ref{alg:online_shapley} converges to also changes from slot to slot. We measure this change by the drift $d_t := \|\boldsymbol{\psi}^*_{t+1} - \boldsymbol{\psi}^*_t\|$. The drift is large, for example, when a vehicle enters the field of view of a CAV that was previously unhelpful, causing its global IV to increase abruptly. Corollary~\ref{cor:adaptive-tracking} extends Theorem~\ref{thm:fixed-convergence} to this time-varying case.
\begin{corollary}\label{cor:adaptive-tracking}
Suppose that Assumptions~1, 2, and 3 hold with a time-varying sampling distribution $q_t$, and that $\alpha_t \kappa \le 1/2$ for every $t$. Let $R_t^2 := \mathbb{E}\big[\|\boldsymbol{\psi}_t - \boldsymbol{\psi}^*_t\|^2\big]$ and $\Lambda_{s,T} := \prod_{t=s}^{T-1}(1 - \alpha_t\mu/2)$, with $\Lambda_{T,T} := 1$. Then,
\begin{equation}
R_T^2 \le \Lambda_{0,T}R_0^2 + \sum_{t=0}^{T-1}\Lambda_{t+1,T}\bigg(3\alpha_t^2\sigma^2 + \frac{2\,\mathbb{E}[d_t^2]}{\alpha_t\mu}\bigg).
\end{equation}
For constant $\alpha_t \equiv \alpha$ and $\mathbb{E}[d_t^2] \le d^2$, this gives $\limsup_{t\to\infty} R_t^2 \le \frac{6\alpha\sigma^2}{\mu} + \frac{4d^2}{\alpha^2\mu^2}$.
\end{corollary}
\begin{proof}
    See Appendix \ref{app:colproof}
    \vspace{-2mm}
\end{proof}
\par Corollary~\ref{cor:adaptive-tracking} shows that, when the IV of the agents varies over time, e.g., due to changes in the scenes or due to a shift in the scheduling policy which causes $\boldsymbol{\psi}^{\ast}$ to change, the error in the IV estimate contains an additional term $4d^2/(\alpha^2\mu^2)$. Unlike the noise term, this term decreases with $\alpha$, since a larger learning rate allows the estimate to follow changes faster. Therefore, the BS should select $\alpha$ according to how quickly the value of the agents changes. A larger $\alpha$ suits scenarios in which the relevance of agents changes often, e.g., dense traffic with frequent occlusions, whereas a smaller $\alpha$ suits slowly changing scenes.
Theorem~\ref{thm:fixed-convergence} and Corollary~\ref{cor:adaptive-tracking} both require $\mu > 0$ (Assumption \ref{assump:A3}, i.e., every agent must be included in the scheduled coalitions with nonzero probability. If the BS never schedules agent $n$, no observed utility carries information about $n$. In that case, $\psi_{n,t}$ is never updated and $\boldsymbol{G}_t$ is singular. This occurs in practice when, e.g., a greedy scheduler that always selects the agents with the highest current global IV never schedules an agent whose IV was underestimated. Proposition~\ref{prop:exploration} shows that $\epsilon$-greedy exploration prevents this.

\begin{proposition}\label{prop:exploration}
Suppose that, at every slot, the BS schedules a coalition drawn uniformly at random from the $2^N-2$ proper non-empty coalitions with probability $\epsilon$, and follows its own scheduling policy otherwise, i.e., $q_t = \epsilon\, q_{\mathrm{bal}} + (1-\epsilon)\, q_{\mathrm{sch},t}$, where $q_{\mathrm{sch},t}$ may be degenerate. Let $\boldsymbol{G}_{\mathrm{bal}}$ and $\boldsymbol{G}_{\mathrm{sch},t}$ denote the matrices in \eqref{eq:G-def} and \eqref{eq:b-def} induced by $q_{\mathrm{bal}}$ and $q_{\mathrm{sch},t}$, respectively. Then,
\begin{equation}
\lambda_{\min}(\boldsymbol{G}_t) \ge \frac{\epsilon(N-1)}{N(2^N-2)} + (1-\epsilon)\,\lambda_{\min}(\boldsymbol{G}_{\mathrm{sch},t}),
\end{equation}
so that $\lambda_{\min}(\boldsymbol{G}_t) > 0$ for every $\epsilon > 0$, even when $\boldsymbol{G}_{\mathrm{sch},t}$ is singular.
\end{proposition}
\vspace{-2mm}
\begin{proof}
    See Appendix \ref{app:proofprop}
    \vspace{-2mm}
\end{proof}
\par Proposition~\ref{prop:exploration} guarantees that Assumption \ref{assump:A3} holds with $\mu \ge \epsilon(N-1)/[N(2^N-2)]$ for any scheduling policy. Hence, Theorem~\ref{thm:fixed-convergence} and Corollary~\ref{cor:adaptive-tracking} apply to any scheduler policy, and the global IV estimates are always updated. A larger exploration factor $\epsilon$ speeds up convergence, but may not maximize the inference performance. In practice, $\epsilon$ is therefore kept small or decayed over time, as evaluated in Section~\ref{sims}. The $\epsilon$-greedy mechanism also reduces the sampling bias in \eqref{eq:total-error-decomposition}. Since $\boldsymbol{\psi}^{\ast}_t$ is fitted only over the coalitions that the BS schedules, $\boldsymbol{\psi}_t$ differs from $\boldsymbol{\Phi}_t$. Thus, increasing $\epsilon$ makes the scheduling policy uniform, which reduces the sampling bias term and makes the estimates from Algorithm \ref{alg:online_shapley} close to the true global IV (Shapley value) \eqref{shapleyeq}. With Algorithm \ref{alg:online_shapley}, the BS obtains an estimate of the global IV of every agent in each slot. Next, we combine the global IV and local IV along with resource constraints to design the IV-aware scheduler.
\vspace{-3mm}

\section{Neuro-Symbolic IV-Aware Scheduling}
While the global IV metric quantifies the marginal contribution of each agent's features to the inference task, the proposed goal-oriented IV-aware scheduler must also account for additional requirements specific to the inference task. 
 In particular, resource allocation in an IV-aware network must jointly account for (i) the IV of received representations, (ii) task-level or context-dependent constraints such as reliability or latency of the inference task, and (iii) the dynamics of the wireless channel. A scalar IV-based objective is thus needed to map these heterogeneous attributes (Shapley importance values and logical constraints) into a single decision-making criterion that the scheduler can optimize in real time.
\par However, defining an objective function for such a scheduler often requires assuming a specific functional form over these attributes, which may vary from one system to another. While one objective function may not be applicable for scheduling in all systems, the scheduler makes decisions over variables with logical and causal dependencies. For example, an agent's LiDAR feature may only be valuable for inference if the camera modality is also scheduled (even if it has a high IV). This \emph{contextual awareness} can be instilled in the network layer by specifying the application requirements as logical rules that are otherwise difficult to express via standard objective functions. A logical rule describes an if-then statement over the scheduling variables and IV, linking task requirements to scheduling. By writing the scheduling requirements as logical rules, the network can  (i) encode conditional and non-additive requirements that standard objective functions cannot capture, (ii) add or remove a requirement by adding or removing a single statement, without redesigning the scheduler, and (iii) identify how each statement affects the resource allocation.
\par To use such statements within a learned scheduler, we adopt a neurosymbolic approach. Neurosymbolic AI combines neural networks, which learn from high-dimensional data, with symbolic reasoning, which operates on explicit variables and logical rules \cite{10148662}. In our scheduler, the symbolic part consists of the interpretable quantities over which the statements are formed (global IV, local IV, channel state, etc.) together with the statements themselves. The neural part is a network that maps these quantities and the received representations to a probability distribution over possible allocations. A purely neural scheduler must learn all requirements from the reward signal, requires extensive interaction to train, and it is impossible to infer why an allocation was produced. A purely symbolic scheduler is interpretable, but the objective function must be tuned for every system, and it is not scalable. A neurosymbolic scheduler keeps the requirements as logical rules and learns their relative importance from data using a neural network, keeping both scalability and interpretability for the scheduler.
\par Specifically, we use a neurosymbolic energy-based model (NeSyEBM)~\cite{10.24963/ijcai.2023/461}, in which each logical rule is converted into a continuous penalty, and the energy function is the weighted sum of these penalties. Here, the energy is a score assigned to each possible allocation and not the energy of the network. The energy function serves as the IV-based scheduling objective. At the beginning of each slot, before any payload is transmitted, the BS computes the energy of every possible allocation $\boldsymbol{r}_t \in \Omega$ and selects the allocation with the lowest energy, i.e., the allocation that best satisfies the rules. This objective complements the quantities global IV and local IV. 
The local and global IV represent the value of each agent's data on its own and relative to the other agents, respectively. The energy combines these per-agent values with the symbolic context (channel gains, payload sizes, and RB cost) into one score for each allocation. We therefore use the terms IV-based objective and energy interchangeably.
The advantage of this approach is that a conventional objective, e.g., a weighted sum of the rate and IV of each agent, is fixed. In contrast, the energy is built from requirements stated as logical rules, e.g., \emph{if} a large fraction of the RBs is allocated, \emph{then} the scheduled agents must have a high IV, and the importance of each statement is learned from data rather than being fixed, making the objective applicable for different tasks. 
Such conditional requirements are hard to express with a fixed formula, but each can be written as a logical rule over possible allocations.
\par Our formulation differs from NeuPSL~\cite{10.24963/ijcai.2023/461} in three ways. First, in NeuPSL, neural networks predict the truth values of symbolic predicates and are trained with labeled or weakly labeled data. Since we lack ground truth, our scheduling policy directly outputs a distribution over the allocations and is trained to minimize the energy. Second, our decision variable is a discrete RB allocation over the combinatorial set $\Omega$, rather than a vector of continuous truth values obtained by convex optimization. Third, the weights of the adaptive rules are learned by dual ascent on their expected violation under the current policy, without labels.
\vspace{-3.5mm}
\subsection{IV-based Objective}
\vspace{-1.1mm}
At each time slot $t$, the BS selects an RB allocation $\boldsymbol{r}_t \in \Omega$, where each $r_{n,t} \in \boldsymbol{r}_t$ represents the number of resources allocated.
The system state comprises the neural features $\hat{z}_t$ received from the agents and the symbolic context $s_t = [\boldsymbol{\psi}_t, \boldsymbol{g}_t, \boldsymbol{p}_t, \boldsymbol{\ell}_t]$, where $\boldsymbol{\psi}_t$ is the global IV (Shapley) estimate from Algorithm \ref{alg:online_shapley}, $\boldsymbol{g}_t$ is the vector of instantaneous channel gains, $\boldsymbol{p}_t$ is the vector of the size of the transmitted payloads, and $\boldsymbol{l}_t$ is the vector of agent-reported local IV score. To find the optimal schedule within the feasible set, we map the scheduling requirements for the inference task to an energy function $E(.)$. 
Low-energy allocations satisfy the established logical rules for IV-aware scheduling, while high-energy allocations indicate constraint violations or suboptimal performance. In the wireless system, the rules encode three types of scheduling requirements, which can be updated according to the requirements of the inference task, defined in Appendix~\ref{potentials}. The IV-based statements require that RBs are given to agents whose data is valuable, i.e., agents with a high global or local IV. The reliability rule requires that an agent with a high global IV and a good channel receives enough RBs to deliver its payload within $D_{\textrm{max}}$. The resource rule penalizes the total number of allocated RBs, so that the BS saves bandwidth when additional RBs would not improve the inference. The energy penalty therefore controls the network through these penalties, defined using global IV and local IV along with other wireless parameters. For example, an allocation that gives a high-IV CAV fewer RBs than needed to meet the deadline violates the reliability rule and receives a high energy, so the BS avoids it. Similarly, adding a CAV with a low global IV lowers the average IV of the scheduled agents and adds RB cost, so the allocation without that CAV has lower energy. The learned weights $\boldsymbol{w}_{\mathrm{sy}}$ set how these requirements are traded off.
 Ultimately, the BS obtains the optimal scheduling of the agents by minimizing the constraint-driven energy function,
\begin{equation}\label{alloc}
    \boldsymbol{r}_t^{\ast} = \argmin_{\boldsymbol{r}_t \in \Omega} E(\boldsymbol{r}_t, \boldsymbol{s}_t, \boldsymbol{\hat{z}}_t; \boldsymbol{w}_{\textrm{sy}}).
\end{equation}
Here, the notation $E(.; w)$ separates the variables from the parameters $\boldsymbol{w}_{\textrm{sy}}$. The arguments before the semicolon change at every slot, i.e., the allocation and the current inputs, whereas the rule weights $\boldsymbol{w}_{\mathrm{sy}}$ after the semicolon are learned offline and kept fixed during deployment.The energy function $E$ combines $K$ logical potentials $\phi(\boldsymbol{r}_t,\boldsymbol{s}_t,\hat{\boldsymbol{z}}_t)=[\phi_1,\dots,\phi_K]^\top$, which form the symbolic component of the model. The parameter $\boldsymbol{w}_{sy}$ represents the learnable weights associated with the symbolic component of the model, which enforce the task-oriented logical rules. The energy function $E$ is defined as the sum of weighted symbolic logical potentials as 
\begin{equation}\label{energyfn}
    E(\boldsymbol{r}_t, \boldsymbol{s}_t, \boldsymbol{\hat{z}}_t; \boldsymbol{w}_{\textrm{sy}})= \boldsymbol{w}_{sy}^T \boldsymbol{\phi}(\boldsymbol{r}_t, \boldsymbol{s}_t, \boldsymbol{\hat{z}}_t).
\end{equation}
Each potential $\phi_k$ encodes a logical requirement that the scheduler must adhere to. Since logical rules can make the energy function non-differentiable, the purpose of the logical potentials is to relax the non-differentiable logic using a hinge-loss formulation based on Lukasiewicz continuous-valued logical semantics, \cite{10.24963/ijcai.2023/461}. 
\begin{equation} \label{hinge}
\phi_k(\boldsymbol{r}_t, \boldsymbol{s}_t, \boldsymbol{\hat{z}}_t) = \max\left( c_k(\boldsymbol{r}_t, \boldsymbol{s}_t, \boldsymbol{\hat{z}}_t)), 0 \right)^{\rho}, \end{equation}
where $c_k(.)$ is a linear function and $\rho \in \{1,2\}$. The hinge-loss formulation is intuitive; if the scheduler perfectly satisfies statement $k$, the potential evaluates to 0 (no energy penalty). If the statement is violated, the energy penalty grows proportionally to the severity of the violation. The wireless constraints enter the logical potentials $\phi(.)$ through the required SINR, $\gamma_{n,t}$, deadline $D_{\textrm{max}}$ and the budget $M$ and the task knowledge enters through the global and local IV. Because the $K$ logical potentials are defined over different quantities like SINR, IV, etc., each potential is rescaled as,
\begin{equation}
    \tilde\phi_k(r) = \begin{cases}
\dfrac{\phi_k(r) - \phi_k^{\min}}{\phi_k^{\max} - \phi_k^{\min} + \varepsilon}, & \phi_k^{\max}-\phi_k^{\min} > \varepsilon,\\[4pt]
0, & \text{otherwise},
\end{cases}
\label{eq:normalized-phi}
\end{equation}
where $\phi_k^{\min}=\min_{r'\in\Omega}\phi_k(r')$, and $\phi_k^{\max}=\max_{r'\in\Omega}\phi_k(r')$ and $\varepsilon$ is a small positive constant. The min-max normalization makes the learned weights $\boldsymbol{w}_{\textrm{sy}}$comparabale across logical rules.
To summarize, the RB allocation is obtained by minimizing an energy function that combines the global IV, the local IV, the channel gains, and the RB cost through weighted logical rules that define scheduling requirements. 
However, minimizing this energy requires the rule weights $\boldsymbol{w}_{\mathrm{sy}}$, which determine the priority of each rule and are not known a priori. Moreover, evaluating $E$ for every candidate allocation within a slot becomes computationally prohibitive as $|\Omega|$ grows combinatorially with $N$ and $M$. In the next section, we present a solution to address both challenges.
\vspace{-5mm}
\subsection{Unsupervised Learning of the NeSyEBM Scheduler}
\vspace{-1.5mm}
To obtain the optimal resource allocation in \eqref{alloc}, the BS needs to learn the energy function $E(.)$. 
Since the scheduling problem lacks ground-truth labels, we formulate the learning task (learning the parameters) as an unsupervised energy minimization problem. The energy function $E$ serves as an intrinsic supervision signal, guiding the scheduler toward allocations that satisfy the logical constraints (i.e., low-energy regions). As scheduling decisions must be produced within the sub-millisecond duration of a cellular network's time slot, enumerating $\Omega$ as $N$ and $M$ grow becomes computationally prohibitive. To address this, we employ amortized inference, i.e., rather than minimizing $E$ directly at every time slot, we learn a neural network $g_{\textrm{nn}}(\cdot; \boldsymbol{w}_{\textrm{nn}})$ with weights $\boldsymbol{w}_{\textrm{nn}}$ that maps $(\boldsymbol{s}_t, \hat{\boldsymbol{z}}_t)$ directly to a probability distribution over low-energy allocations. Thus, the goal is to learn a scheduler ($\pi_{nn}$) that generates allocations in low-energy regions of the state space (satisfying logic statements) without requiring supervision. 
Essentially, the neural policy amortizes the computational cost of minimizing the energy function every time slot via training. We model this policy as a categorical distribution parameterized by $\boldsymbol{w}_{nn}$ over $\Omega$. 
\begin{equation}\label{eq:policy}
   \Tilde{\boldsymbol{r}}_{t}\! \sim \! \pi_{nn}(\cdot\!\mid\!\boldsymbol{s}_t, \hat{\boldsymbol{z}}_t; \boldsymbol{w}_{nn}))\! =\! \textrm{Cat(Softmax(}g_{nn}(\boldsymbol{\hat{z}}_t, \boldsymbol{s}_t);\!\boldsymbol{w}_{nn}\!).
\end{equation}
The parameters $\boldsymbol{w}_{sy}$ and $\boldsymbol{w}_{nn}$ parameterize the NeSyEBM based scheduler.
The symbolic weights are parameterized as $\boldsymbol{w}_{sy} = K \cdot \mathrm{Softmax}(\zeta)$, where $\zeta \in \mathbb{R}^K$ is a learnable logit vector. This ensures $\boldsymbol{w}_{sy} > 0$ and bounds the weights to a limited scale. We treat $\boldsymbol{w}_{sy}$ as a deterministic function of $\zeta$ and learn it via constrained (Lagrangian dual-ascent) optimization on the expected violation of each rule. Concretely, since $\boldsymbol{r}_t$ is drawn stochastically from $\pi_{nn}$, let $\bar{\phi}_k := \mathbb{E}_{\pi_{nn}}[\tilde{\phi_k(\boldsymbol{r}_t})] = \sum_{\boldsymbol{r} \in \Omega} \pi_{nn}(\boldsymbol{r})\tilde{\phi_k(\boldsymbol{r}})$ denote the expected potential for rule $k$ under the current policy and $\bar{\boldsymbol{\phi}} := [\bar\phi_1, \dots, \bar\phi_K]^\top$. Since
$\boldsymbol{w}_{sy}=K\cdot\mathrm{Softmax}(\zeta)$ is a deterministic, differentiable function of $\zeta$, updating $\zeta$ directly updates $\boldsymbol{w}_{sy}$ through the softmax by the chain rule. The symbolic logits are updated as,
\begin{equation}\label{dual-ascent}
    \zeta \leftarrow \zeta +  \beta_{sy}\,\nabla_\zeta\Big(\boldsymbol{w}_{sy}^\top\bar{\boldsymbol{\phi}}\Big).
\end{equation}
With the energy landscape adapting via \eqref{dual-ascent}, the neural policy is trained to minimize the expected energy as
\begin{equation}
    \min_{\boldsymbol{w}_{nn}} \; \mathbb{E}_{\pi_{nn}}\big[E(\boldsymbol{r}_t,\boldsymbol{s}_t,\hat{\boldsymbol{z}}_t;\boldsymbol{w}_{sy})\big] + \lambda\lVert \boldsymbol{w}_{nn}\rVert_2^2,
\label{eq:policy-obj}
\end{equation}
The gradient is obtained by backpropagating directly through the policy $\pi_{nn}$,
\begin{equation}
    \nabla_{\boldsymbol{w}_{nn}}\mathbb{E}_{\pi_{nn}}[E]\! = \!\sum_{\boldsymbol{r}\in\Omega} E(\boldsymbol{r},\boldsymbol{s}_t,\hat{\boldsymbol{z}}_t;\boldsymbol{w}_{sy})\,\nabla_{\boldsymbol{w}_{nn}}\!\pi_{nn}(\boldsymbol{r}\mid \boldsymbol{s}_t,\hat{\boldsymbol{z}}_t),
\label{eq:exact-grad}
\end{equation}
  By utilizing \eqref{eq:exact-grad}, we can iteratively learn the parameters to define the energy function using the objective in \eqref{energyfn}. Once the offline training is complete, the BS can utilize the learned policy $\pi_{nn}$ to obtain the allocations without the computational complexity of minimizing the energy function every time step in \eqref{alloc}. We now discuss how the proposed framework addresses problem \eqref{pr} in Section \ref{sys2}. Problem \eqref{pr} cannot be solved directly, since $U_t(\mathcal{S})$ is observed only after the scheduled agents transmit and the inference is executed. The proposed framework therefore reformulates $U_t$ with quantities that are available before transmission, such as the local IV, the global IV estimates, the channel gains, and payload sizes that determine whether an agent's data can be delivered within $D_{\textrm{max}}$, i.e., whether the agent belongs to $\mathcal{S}(\boldsymbol{r}_t)$. The energy combines these quantities through the scheduling rules, so that a low-energy allocation assigns RBs to agents whose data is informative, non-redundant, and deliverable within the deadline, while using few RBs. The training in \eqref{dual-ascent}--\eqref{eq:exact-grad} learns the priority of each rule from data. After each slot, the observed utility $U_t(\mathcal{S}_t)$ is used by Algorithm \ref{alg:online_shapley} to update the global IV for the next slot. 
  \vspace{-3mm}
\section{Simulation Results and Analysis}\label{sims}
\begin{table}[t]
\centering
\caption{Wireless simulation parameters.}
\label{tab:wireless-params}
\begin{tabular}{lc}
\toprule
Parameter & Value \\
\midrule
Carrier frequency & 5.9 GHz \\
Bandwidth & 20 MHz \\
Transmit power & 23 dBm \\
Noise floor & $-174$ dBm \\
RB budget $M$ & 20 \\
Deadline $D_{\textrm{max}}$ & 10-100 ms \\
Ray-tracing max depth & 3 \\
Ray-tracing samples & $10^5$ \\
\bottomrule
\end{tabular}
\end{table}
\begin{figure}[t]
  \centering
  \begin{subfigure}[b]{0.48\columnwidth}
    \centering
    \includegraphics[width=\textwidth]{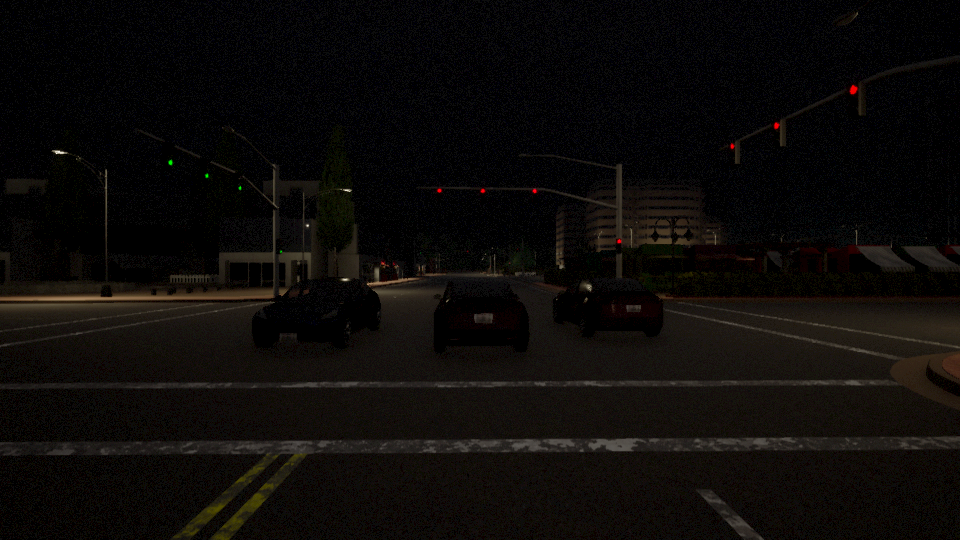}
    \caption{Front camera (Night)}
    \label{fig:sub1}
  \end{subfigure}
  \hfill
  \begin{subfigure}[b]{0.48\columnwidth}
    \centering
    \includegraphics[width=\textwidth]{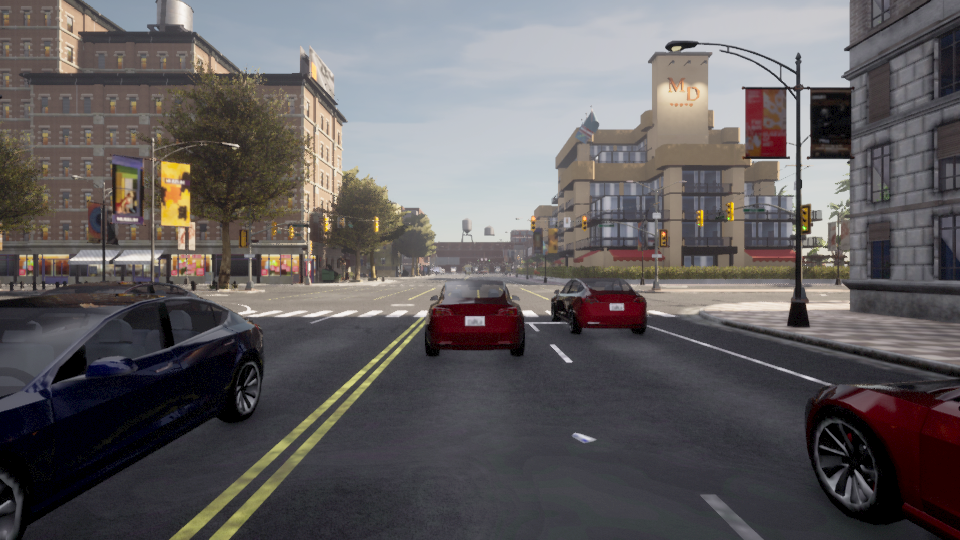}
    \caption{Front camera (Town10)}
    \label{fig:sub2}
  \end{subfigure}
    \vfill
  \begin{subfigure}[b]{0.48\columnwidth}
    \centering
    \includegraphics[width=\textwidth]{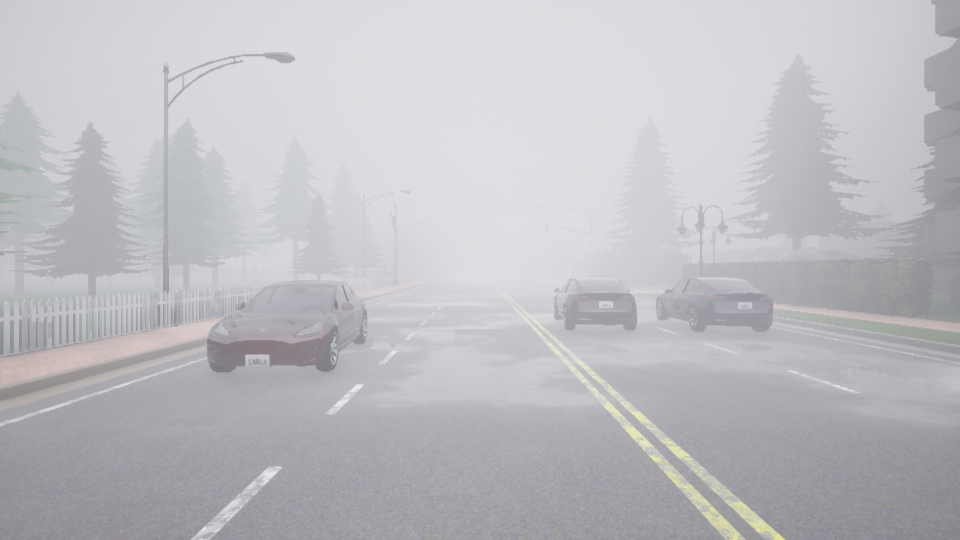}
    \caption{Rear camera (Fog)}
    \label{fig:sub3}
  \end{subfigure}
  \hfill
  \begin{subfigure}[b]{0.48\columnwidth}
    \centering
    \includegraphics[width=\textwidth]{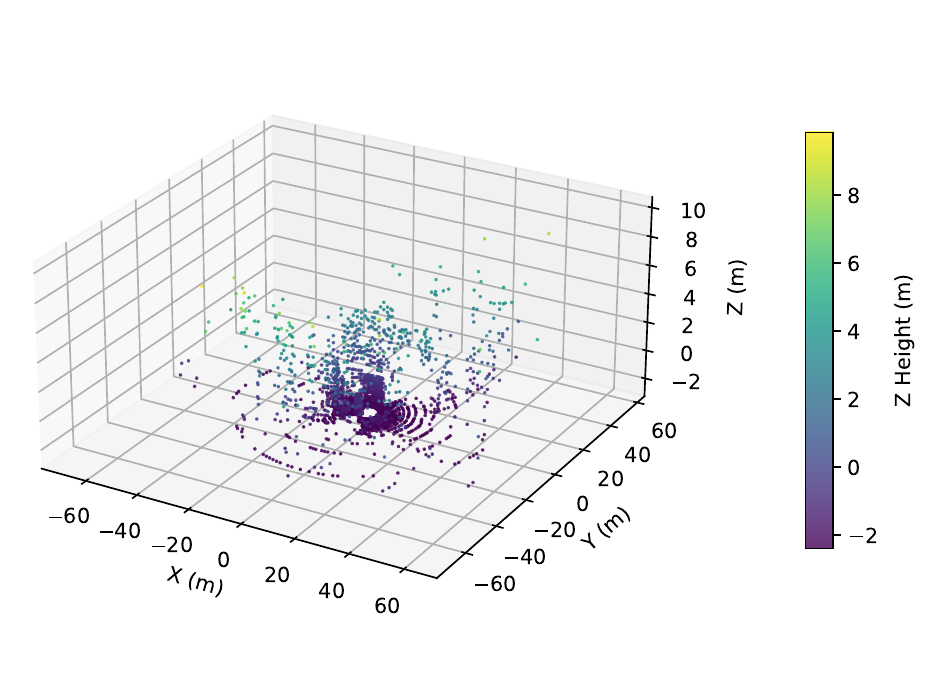}
    \caption{Point cloud visualization}
    \label{fig:sub4}
  \end{subfigure}

  \caption{Examples of RGB image/LiDAR data generated using CARLA.}
  \label{fig:grid_2x2}
 \vspace{-4mm}
\end{figure}
\begin{table*}[htbp!]
\setlength{\tabcolsep}{4pt}
\renewcommand{\arraystretch}{0.95}
\footnotesize
\centering
\caption{Detection F1 on Town05 and Town10 at $D_{\max}=10$~ms and $D_{\max}=100$~ms. }
\label{tab:main-f1}
\begin{tabular}{lcccccccc}
\toprule
 & \multicolumn{4}{c}{$D_{\max}=10$~ms} & \multicolumn{4}{c}{$D_{\max}=100$~ms} \\
\cmidrule(lr){2-5} \cmidrule(lr){6-9}
 & \multicolumn{2}{c}{Town05} & \multicolumn{2}{c}{Town10} & \multicolumn{2}{c}{Town05} & \multicolumn{2}{c}{Town10} \\
\cmidrule(lr){2-3} \cmidrule(lr){4-5} \cmidrule(lr){6-7} \cmidrule(lr){8-9}
Scheduler & Val. & Test & Val. & Test & Val. & Test & Val. & Test \\
\midrule
TaskOnly & 0.616 & 0.644 & 0.648 & 0.663 & 0.616 & 0.644 & 0.648 & 0.663 \\
AoI & 0.711 & 0.776 & 0.699 & 0.699 & 0.711 & 0.821 & 0.725 & 0.714 \\
VoI & 0.804 & \textbf{0.840} & 0.685 & 0.714 & 0.818 & 0.856 & 0.714 & 0.724 \\
MaxRate & 0.717 & 0.771 & 0.649 & 0.677 & 0.717 & 0.771 & 0.649 & 0.677 \\
Proposed NeSyEBM & \textbf{0.844} & 0.829 & \textbf{0.748} & \textbf{0.729} & \textbf{0.882} & \textbf{0.898} & \textbf{0.794} & \textbf{0.780} \\
\midrule
FullFusion & 0.874 & 0.916 & 0.804 & 0.780 & 0.874 & 0.916 & 0.804 & 0.780 \\
\bottomrule
\end{tabular}
\\[2pt]
\end{table*}
\vspace{-1mm}
We evaluate the proposed scheduler in a cooperative perception scenario, a key 5G V2X use case in which vehicles share sensor data to extend their perception beyond their own field of view~\cite{3gpp22186}. Three CAVs with heterogeneous sensors transmit their observations to the BS, which performs cooperative object detection and sends the detections to an ego vehicle whose view is occluded. Vehicle trajectories and sensor ground-truth are generated with CARLA~\cite{Dosovitskiy17}, and the wireless channels are obtained via ray tracing with Sionna RT~\cite{sionna}.  
We evaluate the overall cooperative object detection accuracy measured by the F1 score at an intersection-over-union (IoU) threshold of 0.5, i.e., if the overlap between a bounding box and a ground-truth box, divided by the area of their union, exceeds 0.5. For Town05, we additionally evaluate a target detection task, where we measure whether the ego successfully detects specific ground-truth targets present in the frame with the detection results from the scheduled CAVs. We evaluate these tasks on the held-out validation and test scenario sets that do not occur in the training set. We also evaluate a subset of scenarios under varied weather/lighting conditions (clear day, night, fog) to assess robustness to domain shift. The wireless simulation parameters in Table \ref{tab:wireless-params} follow 3GPP V2X evaluation assumptions \cite{3gpp37885}.

\vspace{-3mm}
\subsection{Implementation Details}
\par \textbf{Dataset Generation.} Each scenario is generated in CARLA along a fixed road geometry, with the ego, three CAVs, and target vehicles placed at longitudinal offsets ($-18$ to $18$\,m in $4$\,m steps) with small random jitter applied to the positions and heading so that the scenarios share a layout but have sufficient variation in placement. The ego and CAV3 carry four cameras (front, left, right, rear) and a single LiDAR sensor; CAV1 carries the same four cameras with no LiDAR; CAV2 carries a single LiDAR sensor with no cameras. We generate 25 scenarios for the training set in Town05, and 8 sets each for the validation and test sets. Each scenario contains 200 frames, where each frame captures RGB views, LiDAR sweep, and GPS poses for every CAV, target, and the ego vehicle. The Town10 dataset is generated using the same pipeline with 20 scenarios in the training set and 5 scenarios each for the validation and test sets. For Town05, we create an additional dataset where we vary the weather and lighting conditions, which has 5 scenarios in the training set and 2 scenarios each for the validation and test sets. Fig. \ref{fig:grid_2x2} shows examples of data generated for each CAV. 
\par \textbf{Ray-traced channel simulation. } For every frame, we obtain $P_{n,t}$ by ray-tracing over each town's 3D geometry, using Sionna RT. Each town's CARLA map assets are converted from their native FBX format into per-category OBJ meshes using Blender, then assembled into a Mitsuba/Sionna XML scene, with an approximate radio material assigned to each mesh category (e.g., building facade, road surfaces, vegetation). For every frame, a Sionna RT transmitter is placed at every CAV's CARLA pose and a receiver at the location of the BS to compute the path gains of the received signals from the CAVs at the BS. Thus, every frame uses a realistic channel realization consistent with the CAV's pose and the town's geometry.
\par \textbf{Perception Model. } We use the SiMO model \cite{wen2026simosinglemodalityoperablemultimodalcollaborative} for cooperative perception at the BS. SiMO is well suited to this setup since CAV1, CAV2, and CAV3 contribute camera-only, LiDAR-only, and combined camera-LiDAR features, respectively, and, hence, the set of modalities present in a given frame changes with whichever CAV subset is scheduled. Standard multimodal fusion architectures assume a fixed, consistent modality set and can produce semantic inconsistency when that assumption breaks. SiMO's Length Adaptive Multimodal fusion (LAMMA) adaptively fuses whichever modality features are present while maintaining semantic space consistency. It can automatically route single-modality frames (CAV1 or CAV2 alone) to the corresponding unimodal branches. The detections from all three branches are fused into a shared bird's-eye-view (BEV) representation. We fine-tune the pretrained SiMO model on our generated Town05 and Town10 datasets to improve prediction accuracy. 
\par \textbf{NeSyEBM training.} The scheduler is trained to minimize the energy \eqref{alloc} in an unsupervised manner. The symbolic weights $\boldsymbol{w}_{sy}$ are parameterized via mean-one softmax and updated by Lagrangian dual ascent ($\beta_{sy}=0.05$), and neural policy weights $\boldsymbol{w}_{nn}$ are trained against the resulting energy with $L_2$
regularization coefficient $10^{-5}$ and gradient clipping at norm $1.0$. The policy network $g_{nn}$ is a single hidden layer of 64 ReLU units, taking the flattened per-agent feature vector as input
and outputs one logit per candidate action in $\Omega$.
Algorithm \ref{alg:online_shapley} uses a learning rate $\alpha=0.05$, over
30 scheduler-training epochs. We define the symbolic potentials used to instantiate NeSyEBM in Appendix \ref{potentials}.
\begin{figure}
    \centering
    \includegraphics[width=\columnwidth]{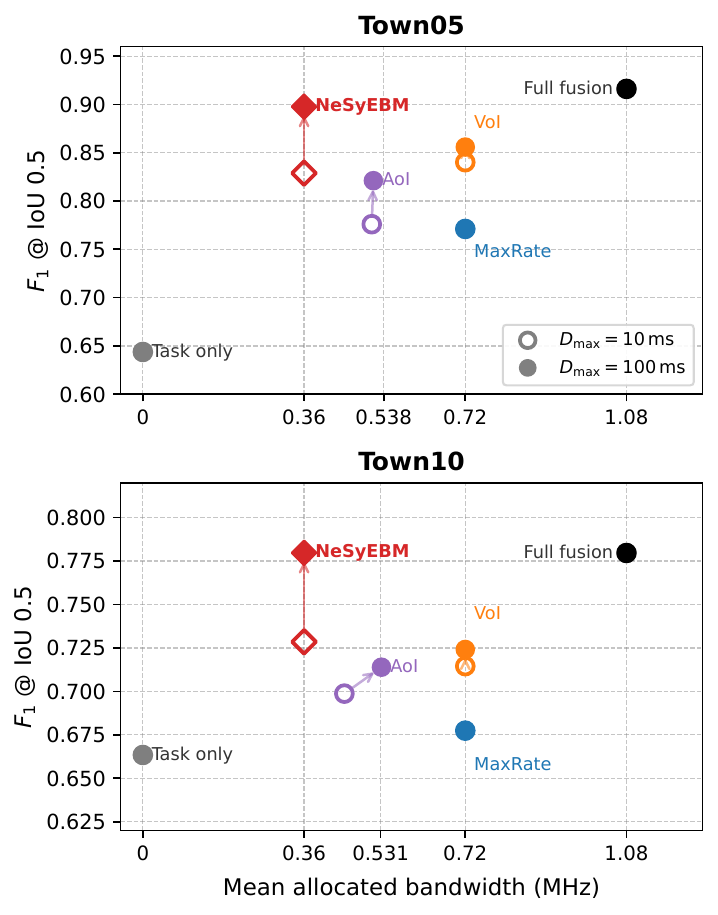}
    \caption{F1 vs. Mean allocated bandwidth.}
    \label{fig:pareto}
    \vspace{-3mm}
\end{figure}
\par \textbf{Baselines.} To show the benefit of cooperative perception in our scenario, we consider two instantiations of SiMO: \emph{TaskOnly}, which refers to the ego vehicle solely relying on its own onboard perception, and \emph{FullFusion}, where we use all the information available from the 3 CAVs regardless of resource constraints. To compare the scheduling performance of the IV-based NeSyEBM scheduler, we compare against \emph{MaxRate}, which schedules the CAVs based on maximizing the instantaneous rate; \emph{AoI}, which schedules the CAVs based on minimizing the AoI; and \emph{VoI}, which schedules the CAV based on a nonlinear function of the AoI and rate.
\vspace{-4mm}
\subsection{Results}
\begin{figure}
    \centering
    \includegraphics[width=\columnwidth]{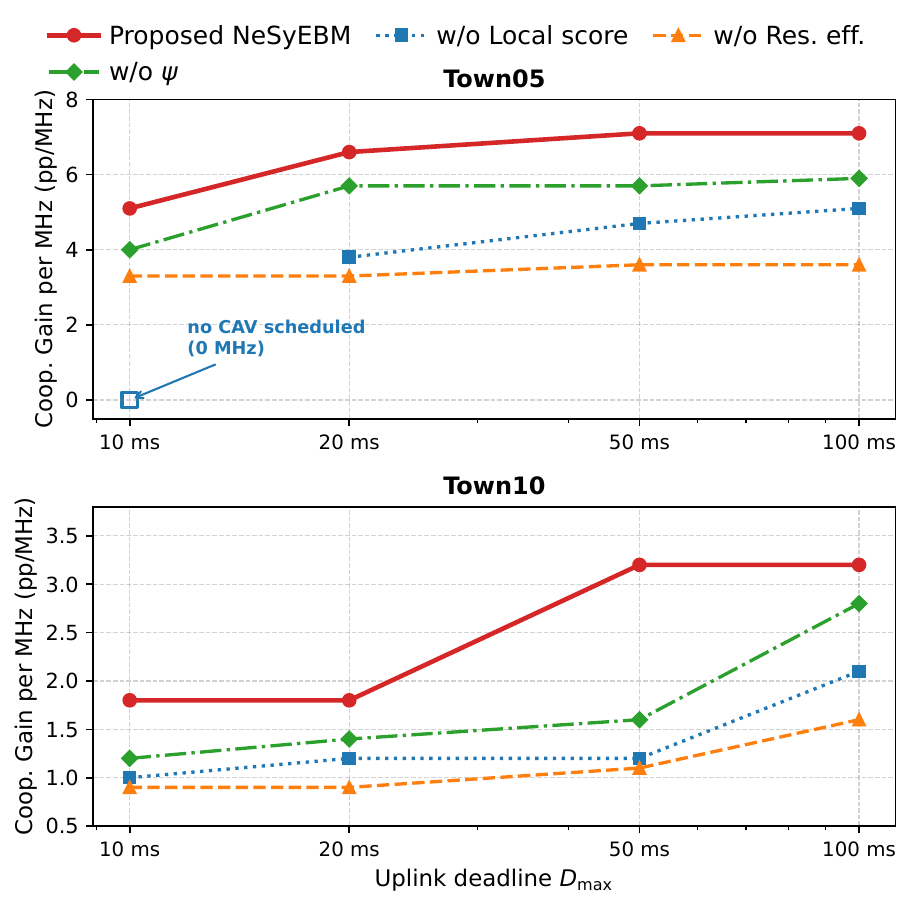}
    \caption{Cooperative gain per MHz of NeSyEBM and its ablations versus the
uplink deadline $D_{\max}$.}
    \label{fig:ablation}
   \vspace{-4mm}
\end{figure}
Table ~\ref{tab:main-f1} presents the achieved F1 score with the baselines across both towns. From Table \ref{tab:main-f1}, we observe that at $D_{\textrm{max}} = 100~\textrm{ms}$ the proposed NeSyEBM scheduler improves the F1 score over MaxRate by 16.4\% in Town05 and 15.1\% in Town10, and over VoI by 4.9\% and 7.7\%, respectively. In fact, the NeSyEBM scheduler comes close to achieving the maximum F1 score referred to as FullFusion, as it represents the budget-unconstrained case. In Fig. \ref{fig:pareto} shows that NeSyEBM achieves this with approximately $50\%$ less bandwidth than MaxRate and VoI. MaxRate and VoI rank CAVs individually, by the wireless channel quality or by age and rate, and therefore do not optimize for bandwidth. In contrast, NeSyEBM uses the global IV to identify the CAV that contributes most to detection given the other CAVs, and the resource-efficiency rule penalizes additional bandwidth allocation when an additional CAV contributes little to the detection accuracy. In our scenarios, a single well-chosen CAV achieves almost the FullFusion accuracy, so a second CAV utilizes additional bandwidth without improving detection. With a shorter deadline $D_{\textrm{max}} = 10~\textrm{ms}$, NeSyEBM achieves the highest F1 score in Town10, while in Town05, its F1 score is $1.3\%$ lower than that of VoI while using half of its bandwidth. This is because a shorter transmission deadline constrains the choices for CAVs for a feasible transmission.
\begin{figure}
    \centering
    \includegraphics[width=\columnwidth]{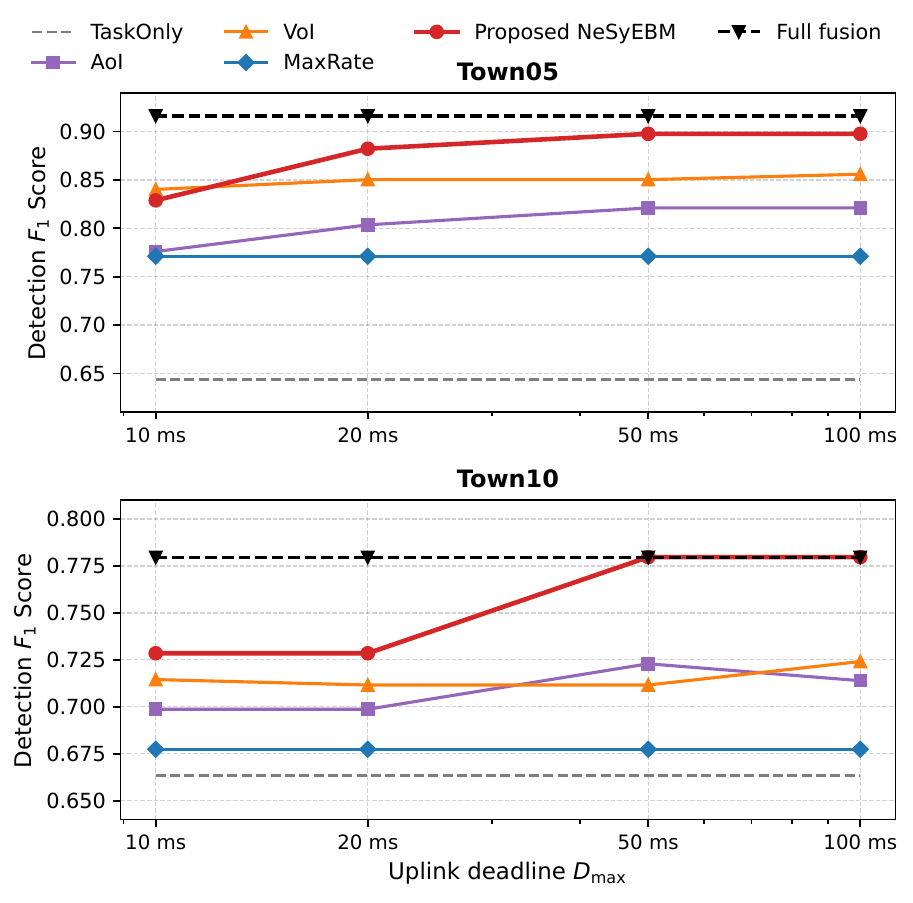}
    \caption{Detection F1 score versus the uplink deadline $D_{\max}$ in Town05 (top) and Town10 (bottom).}
    \label{fig:f1deadline}
   \vspace{-1mm}
\end{figure}
\par In Fig.~\ref{fig:ablation}, we observe the effect of ablating the three most significant logical potentials (or rules) on the bandwidth efficiency. We define the cooperative gain as the increase in F1 score over TaskOnly, in percentage points, per MHz of occupied bandwidth.
This figure allows us to investigate the effect of each logical rule on the proposed scheduler's detection gain per unit of bandwidth.  Removing the resource efficiency rule reduces cooperative gain per MHz by $49.3\%$ for Town05, $50\%$  for Town10 at $D_{\textrm{max}}=100~\textrm{ms}$, and up to $65.6\%$ under tighter deadlines ($D_{\textrm{max}}=50~\textrm{ms}$), while its F1 score remains unchanged. Without this rule, the scheduler reaches a similar accuracy by simply using more RBs to schedule CAVs that contribute little to inference accuracy.
Removing the local IV term has the largest effect on cooperative detection. As observed in Fig. \ref{fig:ablation}, removing the local IV term from the energy function reduces cooperative gain per MHz by $28.2\%$ in Town05 and by $34.4\%$ in Town10 at $D_{\textrm{max}} = 100~\textrm{ms}$, with the reduction increasing to $62.5\%$ at $D_{\textrm{max}} = 50~\textrm{ms}$ in Town10. In Town05, at $D_{\textrm{max}} = 10~\textrm{ms}$, the scheduler allocates no resources (reducing to TaskOnly F1 score). This shows that local IV is essential for the scheduler to have real-time semantic awareness, especially at tighter deadlines. 
Finally, removing the global IV term $\psi$ from the NeSyEBM scheduler reduces cooperative gain per MHz by $16.9\%$ in Town05 and by $12.5\%$ in Town10 at $ 100$ ms. Under tighter deadlines, cooperative gain reduces by up to $21.6\%$ in Town05 (at $ 10$ ms) and by up to $50\%$ in Town10 (at $50$ ms). While the no-$\psi$ NeSyEBM variant is aware of the IV of individual CAVs via the local IV term, it lacks awareness of the IV of the agent with respect to other agents. Under tight latency constraints, we observe that global IV is important for quantifying redundancy among scheduled CAVs, thus ensuring that limited RBs are allocated to CAVs that provide complementary perspectives. 
\begin{figure}
    \centering
    \includegraphics[width=\columnwidth]{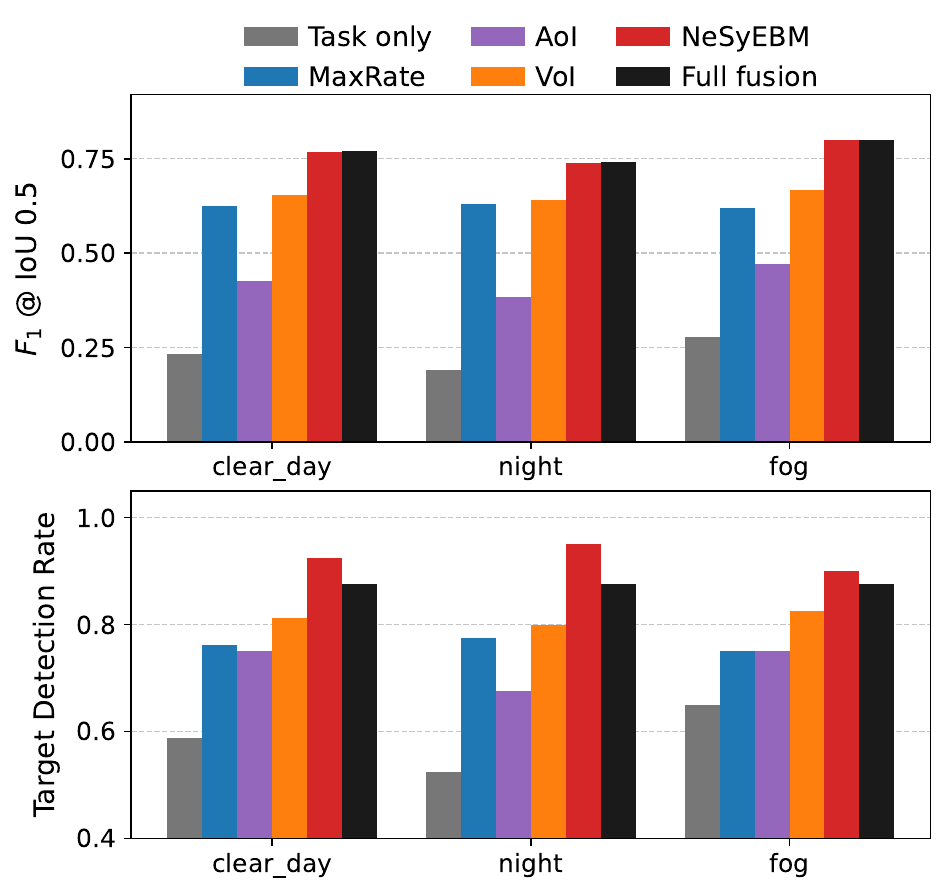}
    \caption{Mean F1 score (Top) and Target detection rate (Bottom) by weather/lighting conditions.}
    \label{fig:weather_lighting}
    \vspace{-3mm}
\end{figure}
\par In Fig. \ref{fig:f1deadline}, we observe the effect of the transmission deadline on the F1 score. We observe that NeSyEBM achieves the highest F1 score at all deadlines in Town10 and for $D_{\textrm{max}} \geq 20~\textrm{ms}$ in Town05. The achieved F1 score improves by $6.4\%$ between 10~ms and 20~ms in Town05 and by $7.0\%$ in Town10, while saturating at $D_{\textrm{max}} = 50~\textrm{ms}$. This is because a longer transmission deadline allows more CAVs to be feasible for uplink transmission, which allows NeSyEBM to allocate resources to high IV CAVs that may require more time due to a larger payload size or a weaker channel. NeSyEBM achieves a higher transmission latency than MaxRate and VoI as it utilizes significantly less bandwidth; however, its 95th percentile latency remains below 42 ms at $D_{\textrm{max}}=100~\textrm{ms}$ and below 4.8 ms at $D_{\textrm{max}}= 10~\textrm{ms}$ for both towns. This is because the NeSyEBM utilizes the available latency margin to improve task performance while reducing bandwidth usage and being well below $D_{\textrm{max}}$. In fact, in case an application requires a different tradeoff between application performance, bandwidth usage, and latency, the NeSyEBM scheduler can be extended with an additional symbolic potential that penalizes the transmission time more, making the scheduler flexible.
\begin{figure}
    \centering
    \includegraphics[width=\linewidth]{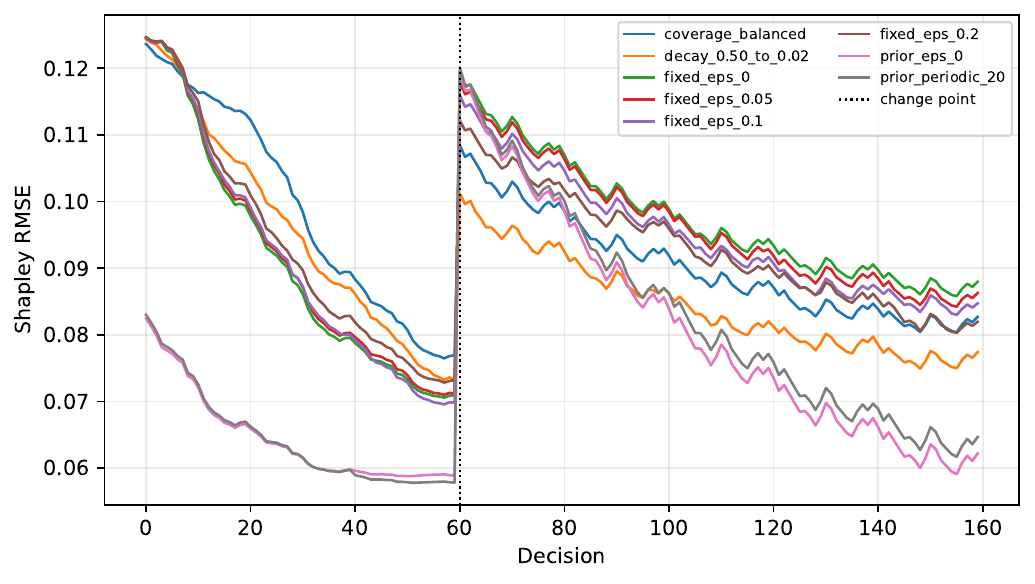}
    \caption{RMSE of Shapley values (global IV) for CAV2 when relevance suddenly shifts at $T=60$.}
    \label{fig:shapley_rmse}
    \vspace{-4mm}
\end{figure}
\par In Fig. \ref{fig:weather_lighting}, we evaluate the robustness of the NeSyEBM scheduler under a domain shift; i.e., the NeSyEBM scheduler is trained only on clear-day scenarios and is evaluated without retraining at night and in fog. Under these conditions, the visibility of the camera sensors is reduced, so the global IV, the local IV, and the relative usefulness of the camera and LiDAR CAVs differ from those seen during training. We compare two tasks on the additional weather/lighting dataset split based on Town05. We can observe that NeSyEBM achieves a higher detection accuracy (F1 score) on all weather conditions (exceeding VoI by $18.5\%$ on a clear day, $15.6\%$ at night, and $21.2\%$ in fog), closely matching FullFusion. This, in turn, shows that the IV metrics capture how the value of each CAV changes with the sensing conditions, which rate- and age-based schedulers cannot observe. We also evaluate a target detection task, where the ego vehicle must detect a specific target (vehicle) which may be critical, e.g., a vehicle occluded by a building at a busy intersection. We observe that NeSyEBM achieves a higher target detection rate than all schedulers in all weather conditions, showing that both the global IV and local IV help identify which CAV is most useful for locating a particular target. Interestingly, we see that NeSyEBM also outperforms FullFusion (by $5.7\%$ on a clear day, $8.6\%$ at night, and $2.9\%$ in fog), with the gap being the smallest in foggy weather conditions. This is because, for the target detection task, fusing every available CAV's data is not guaranteed to be more informative, as compared to aggregate object detection. As multimodal fusion models are highly non-linear, modalities beyond the most informative for the specific target can introduce noise that hurts localization, even if it improves scene-level aggregate F1 score. 
\par To evaluate Algorithm \ref{alg:online_shapley} under non-stationarity, we create a controlled experiment in which the global IV of CAV2 increases abruptly at time step $T=60$, e.g., when a vehicle enters its field of view. Figure~\ref{fig:shapley_rmse} shows the root mean square error (RMSE) between the online global IV and the exact global IV for uniformly initialized ($\boldsymbol{\psi}_{0,n} = \Delta/N$, $\Delta = U(\mathcal{N})-U(\emptyset)$) and a prior-informed initialization (initialized from the exact, offline-computed Shapley values on the training set). Before, $T=60$, prior-informed initialization results in a lower RMSE than uniform initialization. At $T=60$, the error of both prior-informed and uniform (with and without exploration) jumps to $\approx 0.12$, i.e., an abrupt change in IV affects both prior-informed and uniformly-initialized scenarios equally. However, after $T=60$, the error in prior-informed reduces the fastest. Among the schemes with uniform initialization, the reduction in error depends upon the exploration factor. Without exploration $(\epsilon = 0)$, the scheduler rarely selects CAV2, so its global IV is rarely updated, and its error remains the highest. With exploration decaying from $0.5$ to $0.02$, the RMSE reduces to $\approx 0.077$. This result is consistent with Proposition \ref{prop:exploration}, where exploration ensures that every CAV is scheduled often enough for its global IV to be updated, which is important when no prior estimate of global IV is available. \vspace{-3mm}
\section{Conclusion}
\vspace{-1mm}
In this paper, we have proposed an IV-driven scheduling framework for physical AI agents that transmit their observations over a bandwidth-constrained uplink for remote inference. We have proposed a local IV metric, estimated at each agent before transmission, and a global IV metric, computed at the BS using the Shapley value, which captures the contribution of the agent's observations relative to other agents. We have proposed an online stochastic approximation that computes the global IV with $\mathcal{O}(N)$ complexity, and we have shown that the estimation error decreases over time when every agent is scheduled with a nonzero probability. We have then combined the two IV metrics with the channel and the resource constraints as logical rules in a NeSyEBM-based scheduler. Simulation results for a cooperative-perception task have shown that the proposed NeSyEBM-based scheduler achieves around $15\%$ higher F1 score than a MaxRate scheduler and up to $7.7\%$ higher detection accuracy than a VoI scheduler, while using $50\%$ less bandwidth. The ablation studies have shown that removing the local IV or global IV terms reduces the cooperative gain per MHz by up to $62.5\%$ and $50\%$, respectively, under tight deadlines, whereas the resource-efficiency rule halves the bandwidth usage with less than $1\%$ loss in detection accuracy. \vspace{-4mm}
\appendices
\section{Proof of Theorem~\ref{thm:fixed-convergence}}\label{app:proofub}
\begin{proof}
Let $\boldsymbol{v}_t := \boldsymbol{\psi}_t - \boldsymbol{\psi}^*$ denote the estimation error, $\boldsymbol{A}_t :=\omega(\mathcal{S}_t)\boldsymbol{k}_t(\mathcal{S}_t)\boldsymbol{k}_t(\mathcal{S}_t)^\top$ the information matrix of the coalition scheduled at slot $t$, and $\boldsymbol{\nu}_t := \omega(\mathcal{S}_t)\boldsymbol{k}_t(\mathcal{S}_t)\big(\boldsymbol{k}_t(\mathcal{S}_t)^\top\boldsymbol{\psi}^* - \bar{U}_t(\mathcal{S}_t)\big)$ the gradient noise. Substituting the Global IV update gives $\boldsymbol{v}_{t+1} = (\boldsymbol{I} - \alpha_t\boldsymbol{A}_t)\boldsymbol{v}_t - \alpha_t\boldsymbol{\nu}_t$. By the definition of $\boldsymbol{\psi}^*$, $\mathbb{E}[\boldsymbol{\nu}_t \mid \mathcal{F}_t] = \boldsymbol{G}\boldsymbol{\psi}^* - \boldsymbol{b} = \boldsymbol{0}$, and $\mathbb{E}[\boldsymbol{A}_t \mid \mathcal{F}_t] = \boldsymbol{G}$. Expanding the squared norm and taking the conditional expectation gives $\mathbb{E}\big[\|\boldsymbol{v}_{t+1}\|^2 \mid \mathcal{F}_t\big] \le \|\boldsymbol{v}_t\|^2 - 2\alpha_t\boldsymbol{v}_t^\top\boldsymbol{G}\boldsymbol{v}_t + 2\alpha_t^2\,\boldsymbol{v}_t^\top\mathbb{E}[\boldsymbol{A}_t^2 \mid \mathcal{F}_t]\,\boldsymbol{v}_t + 2\alpha_t^2\sigma^2,$ where the last term follows from Assumption \ref{assump:A2}. Since $\boldsymbol{A}_t^2 = \omega(\mathcal{S}_t)\|\boldsymbol{k}_t\|^2\boldsymbol{A}_t \preceq \kappa\boldsymbol{A}_t$ by Assumption \ref{assump:A1}, the third term is at most $2\alpha_t^2 L\,\boldsymbol{v}_t^\top\boldsymbol{G}\boldsymbol{v}_t \le \alpha_t\boldsymbol{v}_t^\top\boldsymbol{G}\boldsymbol{v}_t$, where we used $\alpha_t \kappa \le 1/2$. Using $\boldsymbol{G} \succeq \mu\boldsymbol{I}$ (Assumption \ref{assump:A3}), we obtain the one-step inequality, \vspace{-1mm}
\begin{equation}\label{onestepthm}
\mathbb{E}\big[\|\boldsymbol{v}_{t+1}\|^2 \mid \mathcal{F}_t\big] \le (1-\mu\alpha_t)\|\boldsymbol{v}_t\|^2 + 2\alpha_t^2\sigma^2. 
\end{equation}
Taking the full expectation and applying \eqref{onestepthm} for $t = 0, \dots, T-1$ gives \eqref{eq:finite-time-bound}.
\vspace{-6mm}
\end{proof}
\section{Proof of Corollary \ref{cor:adaptive-tracking}}\label{app:colproof}
\begin{proof}
Let $\boldsymbol{u}_{t+1} := \boldsymbol{\psi}_{t+1} - \boldsymbol{\psi}^*_t$ and $\boldsymbol{\delta}_t := \boldsymbol{\psi}^*_{t+1} - \boldsymbol{\psi}^*_t$, so that $\boldsymbol{\psi}_{t+1} - \boldsymbol{\psi}^*_{t+1} = \boldsymbol{u}_{t+1} - \boldsymbol{\delta}_t$ and $\|\boldsymbol{\delta}_t\| = d_t$. Since $\boldsymbol{u}_{t+1}$ is the error after one update toward the target $\boldsymbol{\psi}^*_t$, \eqref{onestepthm} from the proof of Theorem~\ref{thm:fixed-convergence} gives $\mathbb{E}\|\boldsymbol{u}_{t+1}\|^2 \le (1-\alpha_t\mu)R_t^2 + 2\alpha_t^2\sigma^2$. Let $a_t := \alpha_t\mu$. Since $\mu \le \lambda_{\max}(\boldsymbol{G}_t) \le \kappa$ by Assumption \ref{assump:A1}, we have $a_t \le \alpha_t \kappa \le 1/2$. For any $\chi_t > 0$, Young's inequality gives $\|\boldsymbol{u}_{t+1} - \boldsymbol{\delta}_t\|^2 \le (1+\chi_t)\|\boldsymbol{u}_{t+1}\|^2 + (1+1/\chi_t)\|\boldsymbol{\delta}_t\|^2$, and taking expectations gives $R_{t+1}^2 \le (1+\chi_t)\mathbb{E}\|\boldsymbol{u}_{t+1}\|^2 + (1+1/\chi_t)\mathbb{E}[d_t^2]$. We choose $\chi_t = a_t/[2(1-a_t)]$, for which $(1+\chi_t)(1-a_t) = 1-a_t/2$, $2(1+\chi_t) \le 3$, and $1+1/\chi_t \le 2/a_t$. This gives the one-step inequality,\vspace{-1.5mm}
\begin{equation}\label{one-step}
R_{t+1}^2\!\le \! \Big(1-\frac{\alpha_t\mu}{2}\Big)R_t^2 + 3\alpha_t^2\sigma^2 + \frac{2\,\mathbb{E}[d_t^2]}{\alpha_t\mu}.
\end{equation}
Applying \eqref{one-step} for $t = 0, \dots, T-1$ gives the bound in Corollary~\ref{cor:adaptive-tracking}. For constant $\alpha$ and $\mathbb{E}[d_t^2] \le d^2$, $\Lambda_{s,T} = (1-\alpha\mu/2)^{T-s}$, and summing the geometric series gives $R_T^2 \le (1-\alpha\mu/2)^T R_0^2 + \big(\frac{6\alpha\sigma^2}{\mu} + \frac{4d^2}{\alpha^2\mu^2}\big)\big[1-(1-\alpha\mu/2)^T\big]$. Letting $T \to \infty$ gives the limsup bound.
\end{proof}
\vspace{-6mm}
\section{Proof of Proposition \ref{prop:exploration}}\label{app:proofprop}
\begin{proof}
$\boldsymbol{G}_t$ is a convex combination of $\boldsymbol{G}_{\mathrm{bal}}$ and
$\boldsymbol{G}_{\mathrm{sch},t}$ under the mixture $q_t$, and the minimum
eigenvalue $\lambda_{\min}(\cdot)$ is concave on the cone of positive
semidefinite matrices, giving
$\lambda_{\min}(\boldsymbol{G}_t)\ge\epsilon\lambda_{\min}(\boldsymbol{G}_{\mathrm{bal}})+(1-\epsilon)\lambda_{\min}(\boldsymbol{G}_{\mathrm{sch},t})$.
Since $q_{\mathrm{bal}}$ treats every agent symmetrically,
$\boldsymbol{G}_{\mathrm{bal}}$ takes the form $a\boldsymbol{I}+b\boldsymbol{J}$ for
constants $a,b$ depending on $N$ ($\boldsymbol{J}$ the all-ones matrix), with
eigenvalues $a$ (multiplicity $N-1$) and $a+Nb$ (multiplicity 1), so
$\lambda_{\min}(\boldsymbol{G}_{\mathrm{bal}})=a$. Writing $D,O$ for
$\boldsymbol{G}_{\mathrm{bal}}$'s diagonal and off-diagonal entries
($a=D-O$), evaluating them from \eqref{eq:G-def}, \eqref{eq:b-def} gives
$D=(N-1)H_{N-1}/[N(2^N-2)]$ and $O=[(N-1)H_{N-2}-(N-2)]/[N(2^N-2)]$,
where $H_m:=\sum_{j=1}^m1/j$; simplifying $D-O$ using
$H_{N-1}-H_{N-2}=1/(N-1)$ gives
$\lambda_{\min}(\boldsymbol{G}_{\mathrm{bal}})=D-O=(N-1)/[N(2^N-2)]$.
\end{proof}
\vspace{-6mm}
\section{Potentials defined for Simulations}\label{potentials}
In our simulations, $K=5$. The logits of the potentials are updated by dual ascent \eqref{dual-ascent}. Potentials $\phi_{\textrm{imp}}$ and $\phi_{\textrm{rel}}$ use the exponent $\rho=2$ and all other
potentials use $\rho=1$.\\
\textit{Notation.} For an allocation $\boldsymbol{r}$, let
$\mathcal{S}(\boldsymbol{r})=\{n: r_n>0\}$ be the set of scheduled agents. For a
per-agent quantity $x_n$, $\bar{x}_{\mathcal{S}}$ denotes its mean over
$\mathcal{S}(\boldsymbol{r})$ (zero if $\mathcal{S}(\boldsymbol{r})=\emptyset$),
and $T_{\mathcal{S}}(x)$ denotes the mean of the $|\mathcal{S}(\boldsymbol{r})|$
largest values of $x_n$ over all agents, i.e., the best value achievable with
the same number of scheduled agents. We write $\tilde{x}_n = \max(0,x_n)/\max_m
\max(0,x_m)$ for max-normalization to $[0,1]$ and $[x]_0^1$ for clamping to
$[0,1]$. 
 The IV-based potentials reward or penalize scheduling decisions based on the IV of each agent's observation, independent of resource cost. The importance potential is defined as,
\begin{equation}
\phi_{\mathrm{imp}}(\boldsymbol{r},\hat{\boldsymbol{z}}_t) = \sum_n \Big(\max\big(0,\ (l_{n,t}\widetilde{\psi}_{n,t} - r_n/M\big)\Big)^2,
\label{eq:phi-nov-app}
\vspace{-1mm}
\end{equation}
The reliability potential is defined as,
\begin{equation}
\phi_{\mathrm{rel}}(\boldsymbol{r},\boldsymbol\psi_t,\boldsymbol{g}_t) = \sum_{n=1}^{N} \Big(\max\big(0,\ \widetilde\psi_{n,t}\cdot\widetilde g_{n,t} - \tilde{o}(r_n)\big)\Big)^2,
\label{eq:phi-imp-app}
\vspace{-1mm}
\end{equation}
where $\tilde{o}(r_n) = [\exp(-\gamma_{\mathrm{req}}(r_n)/\gamma_n)]_0^1$ when
$r_n>0$ and $0$ otherwise, with $\gamma_{\mathrm{req}}(r_n) = 2^{L_n/(r_n B D_{\max})}-1$ the SINR required to deliver payload $L_n$ in $r_n$ RBs within $D_{\max}$, and $\gamma_n$ the agent's realized SINR. Although the ray-traced channel is deterministic, this smooth exponential provides a differentiable form for link feasibility.
The global IV potential is defined as,
\begin{equation}
\phi_{\psi}(\boldsymbol{r}) = \max(0,\ T_{\mathcal{S}}(\tilde{\psi}) - \bar{\tilde{\psi}}_{\mathcal{S}}),
\end{equation}
which penalizes allocations whose scheduled agents have a lower mean normalized
global IV than the $|\mathcal{S}(\boldsymbol{r})|$ agents with the highest global
IV. 
The local-IV potential $\phi_{\ell}$ is defined as,
\begin{equation}
\phi_{\ell}(\boldsymbol{r}) = \max(0,\ T_{\mathcal{S}}(\tilde{\ell}) - \bar{\tilde{\ell}}_{\mathcal{S}}),
\end{equation}
the same construction applied to the local-IV score $\ell_{n,t}$.
The resource-efficiency potential is defined as,
\begin{equation}
\phi_{\mathrm{eff}}(\boldsymbol{r}) = \max\big(0,\ \sum_n r_n/M - [\max(\bar{\tilde{\ell}}_{\mathcal{S}},\,
\bar{\psi}_{\mathcal{S}})]_0^1\big),
\end{equation}
which penalizes allocations that use a larger fraction of the RB budget than
justified by the IV of the scheduled agents.
\vspace{-2mm}
{\footnotesize
\def\baselinestretch{0.89}
\bibliographystyle{IEEEtran}
\bibliography{IEEEabrv, semantic_refs}
}
\end{document}